\documentclass[useAMS,usenatbib]{mn2e}

\usepackage{graphicx}

\title[Pseudo-bulges, classical bulges and ellipticals in SDSS]{Structural properties of pseudo-bulges,
classical bulges and elliptical galaxies: an SDSS perspective}
\author[Dimitri A. Gadotti]{Dimitri A. Gadotti\thanks{E-mail: dimitri@mpa-garching.mpg.de}\\
Max-Planck-Institut f\"ur Astrophysik, Karl-Schwarzschild-Str. 1, D-85748
Garching bei M\"unchen, Germany}
\begin{document}


\pagerange{\pageref{firstpage}--\pageref{lastpage}} \pubyear{2008}

\maketitle

\label{firstpage}

\begin{abstract}
We have performed 2D bulge/bar/disc decompositions using $g$, $r$ and $i$-band images of a representative
sample of nearly 1000 galaxies from the Sloan Digital Sky Survey.
We show that the Petrosian concentration index is a better proxy
for bulge-to-total ratio than the global S\'ersic index. We show that pseudo-bulges can be
distinguished from classical bulges as outliers in the Kormendy relation. We provide the structural parameters
and distributions of stellar masses of ellipticals, classical bulges, pseudo-bulges, discs and bars,
and find that 32 per cent of the total stellar mass in massive galaxies in the local universe is contained
in ellipticals, 36 per cent in discs, 25 per cent in classical bulges, 3 per cent in pseudo-bulges
and 4 per cent in bars. Pseudo-bulges are currently undergoing intense star formation activity and populate
the blue cloud of the colour-magnitude diagram. Most (though not all) classical
bulges are quiescent and populate the red sequence of the diagram. Classical bulges follow a correlation
between bulge S\'ersic index and bulge-to-total ratio, while pseudo-bulges do not. In addition, for a fixed
bulge-to-total ratio, pseudo-bulges are less concentrated than classical bulges.
Pseudo-bulges follow a mass-size relation similar
to that followed by bars, and different from that followed by classical bulges.
In the fundamental plane, pseudo-bulges occupy the same locus as discs. While these results point out different
formation processes for classical and pseudo-bulges, we also find a significant overlap in their properties,
indicating that the different processes might happen
concomitantly. Finally, classical bulges and ellipticals follow offset mass-size relations, suggesting
that high-mass bulges might not be simply high-mass ellipticals surrounded by discs.
\end{abstract}


\section{Introduction}
\label{sec:intro}

The central components of disc galaxies have proven deceptively simple. The early view of galaxy bulges
as scaled ellipticals is far from complete. Even the mere definition of a galaxy bulge is still prone to debate.
Yet bulges hold crucial clues to galaxy formation and evolution. Current views discern between bulges formed
through violent processes, such as hierarchical clustering via minor mergers, named classical bulges,
and those formed through longer time-scales, via disc instabilities and secular evolution processes,
named pseudo-bulges \citep[see][for reviews]{WysGilFra97,KorKen04}. Given their dissimilar origins,
it is naturally expected that these different bulge categories should
be structurally distinct.

In fact, previous work have showed evidence that pseudo-bulges, when compared to classical bulges,
tend to show younger stellar populations, kinematics supported by rotation, and less concentrated
surface brightness profiles, similar to those of discs,
which can be parameterised by the corresponding S\'ersic indices
\citep[see e.g.][and references therein]{CarStideZ97,Gaddos01,KorCorBlo06,DroFis07,FisDro08}.
In relatively inclined galaxies, pseudo-bulges are seen to be more flattened than classical bulges, and as
flattened as discs. Some pseudo-bulges display a characteristic box/peanut shape if the galaxy
is viewed edge-on. Such box/peanut bulges are believed to be the inner parts of bars that have evolved
and buckled off the plane of the disc \citep[see e.g.][and references therein]{Ath05b}.
Such evidence is, however, limited. Since such studies brought along new perceptions, proofs of concept,
and because pseudo-bulges are less conspicuous than classical bulges, they had to focus
on a detailed analysis of small samples, often with high resolution data.
For instance, pseudo-bulges often host nuclear bars and/or nuclear spiral arms, which are
difficult to identify without images of high physical spatial resolution, or a
careful analysis.

Ellipticals would be formed through major mergers. The gas content in galaxies, as a dissipative
component and the material to form new stars, is likely to play an important role in minor/major
mergers and in secular evolution, and adds considerable complexity to these views. A more complete picture
of galaxy formation and evolution demands a deep understanding of the consequences of such diverse
processes, as well as the structural properties of the different galaxy components.
It is still necessary to understand better how these different formation processes result in
evolutionary and structural differences, how these differences translate into
observables, and how this framework can be incorporated by theoretical developments.
Questions such as how structurally different ellipticals, classical bulges and pseudo-bulges are,
how different the host galaxies of classical and pseudo-bulges are,
how often disc galaxies host pseudo-bulges as compared to classical bulges, how are their
masses distributed, and in which conditions pseudo-bulges are important, remain to be
properly answered. Time is ripe for more extensive studies on the properties of bulges and ellipticals,
taking advantage of large datasets, such as the Sloan Digital Sky Survey (SDSS), which are becoming more
commonly available.

We have performed a detailed structural analysis of a sample of nearly 1000 galaxies from the SDSS.
Through such a work, we were able to determine S\'ersic indices, scale-lengths and effective
surface brightness for elliptical galaxies, classical bulges and pseudo-bulges, and scale-lengths
and central surface brightness for discs, taking into account the presence of a bar when necessary,
as well as bulge-to-total and disc-to-total luminosity ratios, in three different bandpasses
($g$, $r$ and $i$).
We have also determined colours and stellar masses for each galaxy component separately.
Such structural parameters are explored throughout this paper, combined with several other physical
properties obtained for the SDSS sample of galaxies. (A similar set of parameters were also
obtained for bars, which are investigated in a separate paper.)

\citet{AllDriGra06} and \citet{BenDzaFre07} have recently performed detailed structural analysis
using large datasets. While their samples are substantially larger than ours, our methodology
aims at a more careful analysis.
For instance, we have included a bar component in the models fitted, since, as
shown in \citet{Gad08b}, if the bar is not taken into account in the modelling of barred
galaxies, bulge properties can be significantly affected. We have produced fits in multiple bands
for every galaxy, which have allowed us to estimate the colours of the different components. With
such colours, we were able to directly estimate mass-to-light ratios for the different components,
which results in more accurate stellar mass estimates.
Finally, we have selected a sample of galaxies suitable for such image decompositions, and each
fit was performed and checked individually. All this care assures us that the results from
our analysis are trustworthy to carry the investigations we aimed for.

A common practice in studies on bulges is to define pseudo-bulges as those with a S\'ersic index
lower than a given threshold. However, the use of S\'ersic indices to
distinguish pseudo-bulges from classical bulges is likely to
generate considerable ambiguity, as the dynamical range of this parameter is small, compared
to the uncertainty in individual measurements \citep[see discussion in][]{Gad08b,DurSulBut08}.
Instead, we have used the Kormendy relation \citep{Kor77} to identify pseudo-bulges. Some degree of
structural similarity between classical bulges and ellipticals is borne out by such relation, which
is followed by both kinds of systems. In contrast,
pseudo-bulges, being structurally different, do not follow a similar relation. Therefore, our
identification of pseudo-bulges is not only more reliable but also better physically motivated.
It is also important to point out that, due to the fact that our models include a bar component, the
pseudo-bulges we identify in this study are not those pseudo-bulges associated with
the box/peanut morphology.

This paper is organised as follows. The next section describes the sample selection and properties. Section
\ref{sec:method} describes the methodology implemented to produce the fits. Results are presented in
Sect. \ref{sec:results}, separated in several subsections. We discuss various aspects of our main
findings in Sect. \ref{sec:dis}, including a comparison with previous work. Finally, we summarise our work
and major conclusions in Sect. \ref{sec:sum}. Throughout this work, we have used the following cosmological
parameters: $H_0=75$ km s$^{-1}$ Mpc$^{-1}$, $\Omega_M=0.3$ and $\Omega_\Lambda=0.7$.
Unless noted, photometric measurements refer to the $i$-band. All logarithms are base 10.

\section{Sample selection and properties}
\label{sec:sample}

\begin{figure*}
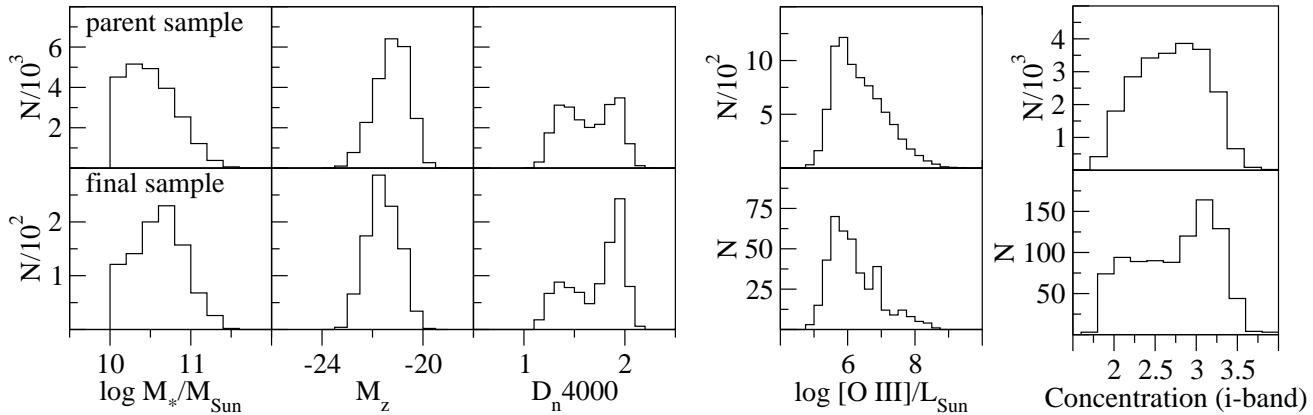

   \centering
   \includegraphics[width=13cm,clip=true,keepaspectratio=true]{scomp_new.eps}
   \includegraphics[width=4.2cm,clip=true,keepaspectratio=true]{c_new_pap.eps}
   \caption{Distributions of galaxy properties for the parent sample (top panels) and the final sample
(bottom panels). From left to right: dust-corrected stellar mass, k-corrected $z$-band absolute
magnitude, D$_n$4000 (corrected for [Ne {\sc iii}] contamination),
extinction-corrected [O {\sc iii}] 5007\AA\ luminosity for AGN, and Petrosian
concentration index. The first three parameters were taken from \citet{KauHecWhi03}, [O {\sc iii}]
luminosities from \citet{KauHecTre03}, and concentration from the SDSS database. The [O {\sc iii}]
luminosity has been used in previous work as an indicator of the accretion rate onto the central
black hole \citep{HecKauBri04}.}
   \label{fig:scomp}
\end{figure*}

We aimed for a sample which is concomitantly suitable for structural analysis based on image
decomposition and a fair representation of the galaxy population in the local universe.
We thus drew the sample from all objects spectroscopically classified as galaxies in the SDSS Data
Release Two (DR2), applying the following criteria. We first selected all galaxies
with redshift in the range $0.02\leq z\leq0.07$. This provides a sample with a
statistically meaningful number of galaxies, whose images have a relatively comparable physical spatial
resolution. Images of closer galaxies have a level
of detail much richer than those of galaxies at $z\approx0.05$, as observed within SDSS.
On the other hand, images of farther galaxies have a physical spatial resolution which is often not adequate
to study their structural properties in some detail.

Secondly, since dwarf galaxies
are not object of our study, we excluded all galaxies with stellar masses below
$10^{10}~{\rm M}_\odot$. Galaxy stellar masses were obtained from \citet{KauHecWhi03}.
At this stage, we have a volume-limited sample, i.e. a sample which includes
all galaxies more massive than $10^{10}~{\rm M}_\odot$ in the volume defined by our redshift
cuts and the DR2 footprint. From now on, we will refer to this sample as our parent sample.

Finally, we chose objects with an axial ratio $b/a\geq0.9$, where $a$ and $b$ are, respectively,
the semi-major and semi-minor axes of the galaxy, taken from the SDSS database (these are measured at the
25 mag arcsec$^{-2}$ isophote in the $g$-band). This criterion assures us that our galaxies are
very close to face-on, meaning that effects from dust attenuation,
as well as projection effects, are minimised, and that
bars can be distinctly seen, avoiding erroneous fits, and leading to a more reliable determination
of the structural parameters and colours of the separate galaxy components. It also
implies we do not need to worry about inclination corrections.
These criteria resulted in a sample of 3375 objects. The images of each of these galaxies were
individually inspected to remove from the sample those galaxies which are either not truly
face-on,\footnote{Axial ratios from SDSS measurements are not always
correct. Galaxies with $b/a$ clearly below $0.9$ were removed from the sample.}
substantially disturbed by a companion or merger, overly faint or irregular, have images too close to the border
of the CCD frame, as well as duplicate entries and those images where the presence of a bright star renders
them useless. We also rejected galaxies with $a<4$ arcsec, as we deemed these galaxies too small for a detailed parametric
image decomposition. This ensured that, for the vast majority of the sample, the effective radius
of the bulge is larger than the PSF HWHM. Hence, reliable conclusions can be drawn even for small bulges
(see discussion further below). Our final sample contains 963 galaxies, of which 407 contain an active nucleus
(AGN -- mostly type 2). AGN classification is taken from \citet{KauHecTre03}.

Most of the galaxies in the final sample have $0.04<z<0.06$. At this distance,
1 arcsec corresponds roughly to 1 kpc. Since the SDSS spectra are taken through a circular aperture of 3 arcsec
in diameter, this means that, for the majority of our galaxies, the spectra obtained represent the central,
bulge-dominated part of the galaxy, with little disc contamination. This will become clearer further below, where the
effective radius of bulges are presented. Since the mean PSF FWHM in SDSS images is $\approx 1.5$
arcsec, the typical physical spatial resolution in the images of our final sample is about 1.5 kpc.
Most of these galaxies also have $0.9\leq b/a<0.95$. In many cases, this means that bulge and disc
have slightly different geometrical properties, such as ellipticity
$\epsilon$ (i.e. $1-b/a$) and position angle. These
differences are helpful when fitting models to the images, since they provide further constraints to the models
of the different galaxy components. Finally, their typical stellar mass is similar to that of the Milky Way
(i.e. $\approx5\times10^{10}$ M$_\odot$ -- \citealt{SomDol01}).

\begin{figure}
   \centering
   \includegraphics[width=7cm,clip=true,keepaspectratio=true]{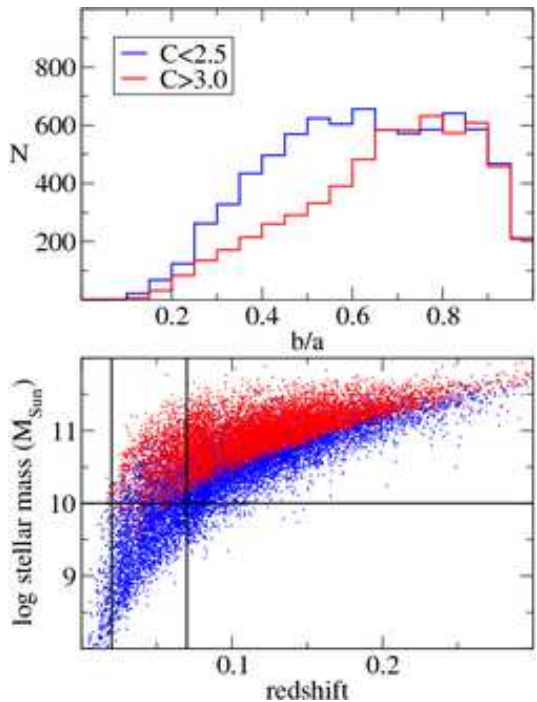}
   \caption{Top: distribution of isophotal axial ratios for galaxies in the parent sample with concentration
index below 2.5 (blue line) and above 3 (red line). Bottom: stellar mass plotted against redshift for all
spectroscopically classified galaxies in SDSS DR2. Black lines indicate our cuts in mass and redshift.
Blue and red dots correspond to galaxies with low and high dust-corrected $z$-band mass-to-light ratios,
respectively.}
   \label{fig:axvol}
\end{figure}

Let us now check whether our final sample is a fair representation of the galaxy population in the parent,
volume-limited sample, and whether our selection criteria to reach the final sample introduce any bias that have
to be accounted for. With this aim, we plot in Fig. \ref{fig:scomp} the distributions of several
relevant properties of the galaxies in both samples. One sees that these distributions are reasonably
similar, in particular those of the AGN [O {\sc iii}] luminosity. However, it is also apparent that
the final sample includes a somewhat larger fraction of massive, brighter,
quiescent and more concentrated galaxies.
The more stringent criterion to reach the final sample from our parent sample is the axial ratio cut.
Thus, we now investigate whether this criterion is the cause of such differences. The top panel of
Fig. \ref{fig:axvol} shows the distributions of axial ratio for galaxies in the parent sample, separated
according to their concentration $C$. From our structural analysis below, we find that most ellipticals,
and few disc galaxies, have $C>3$. Conversely, most disc-dominated galaxies, and few ellipticals, have
$C<2.5$. We thus use these thresholds to separate ellipticals and disc galaxies. It is evident that, due
to their different {\em intrinsic} axial ratios, such galaxies have different distributions of the
observed axial ratio, in such a way that the fraction of elliptical galaxies increases at higher values
of $b/a$. Therefore, it is indeed the axial ratio cut at $b/a\geq0.9$ that produces the differences
seen in Fig. \ref{fig:scomp}. Some of the results presented below depend on the fraction of elliptical
galaxies in the sample, and are thus affected by this selection effect.
We can correct these results for such effect
by comparing the fraction of galaxies with $C>3$ in the parent sample, regardless of axial ratio, to
that of galaxies with both $C>3$ and $b/a\geq0.9$. This provides us with a fairly good indication
of how larger the fraction of elliptical galaxies in our final sample is compared to our parent,
volume-limited sample. It turns out that this difference is a factor of 1.3. Such correction factor
will be applied when necessary with an explicit mention.

Given our axial ratio cut, one might also question whether the results in this work concern galaxies
which are intrinsically more eccentric than average. To check that, we have also selected 30 systems
with $C>2.5$ (i.e. mostly bulge-dominated and elliptical galaxies) and $0.5<b/a<0.7$, which otherwise
comply with all other selection criteria of the final sample, and analyse them in a similar fashion.
One could also ask if we are missing galaxies with very strong bars due to our cut in axial
ratio. This would happen if the outer disc of the galaxy is not detected. We argue that this is not the
case, based on the fact that the strongest bars we find, considering the bar-to-total luminosity
ratio, are about as strong as the strongest ones found in other studies
\citep{SelWil93,ReeWilSel07,Gad08b,DurSulBut08}.

Finally, we also need to check if there is a bias in our parent sample against galaxies close to both
its low mass and high redshift cuts, which could occur due to magnitude limits in the survey.
This is done by plotting galaxy mass against redshift for
all spectroscopically classified galaxies in DR2 in the lower panel
of Fig. \ref{fig:axvol}. Since such a bias might depend on galaxy colour, we plot separately galaxies with
low and high dust-corrected $z$-band mass-to-light ratio \citep[from][]{KauHecWhi03}, using the peak of its
distribution as boundary between the two groups. One sees that there is no clear need for such a volume
correction, i.e. our parent sample includes effectively all galaxies more massive than $10^{10}~{\rm M}_\odot$
in the defined volume.

We have also compared the distributions of other physical parameters, such as stellar surface density and
metalicity, $r$-band light-weighted age, number of neighbors, and the fluxes and equivalent widths of
important emission lines in AGN, derived by \citet{KauHecTre03,KauHecWhi03,KauWhiHec04} and \citet{Gal05},
in both our parent and final samples. These distributions
are generally similar. We thus conclude that, after accounting for the effects introduced with our axial ratio
cut, our final sample is a fair representation of the population of galaxies and AGN in the local universe with
stellar masses larger than $10^{10}~{\rm M}_\odot$. This means that conclusions based on this sample are generally
pertinent to galaxies that satisfy these conditions.

\begin{figure*}
   \centering
   \includegraphics[width=3cm,clip=true,keepaspectratio=true]{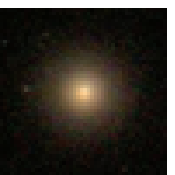}
   \includegraphics[width=3cm,clip=true,keepaspectratio=true]{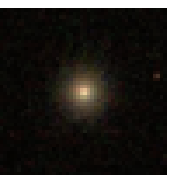}
   \includegraphics[width=3cm,clip=true,keepaspectratio=true]{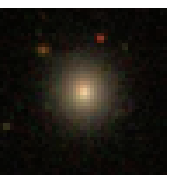}
   \includegraphics[width=3cm,clip=true,keepaspectratio=true]{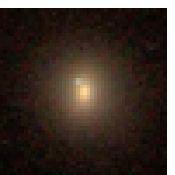}
   \includegraphics[width=3cm,clip=true,keepaspectratio=true]{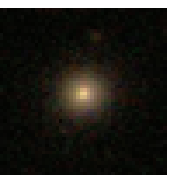}\\
   \includegraphics[width=3cm,clip=true,keepaspectratio=true]{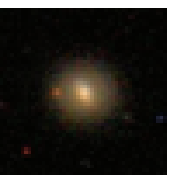}
   \includegraphics[width=3cm,clip=true,keepaspectratio=true]{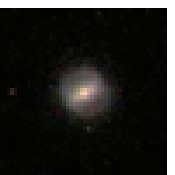}
   \includegraphics[width=3cm,clip=true,keepaspectratio=true]{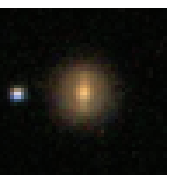}
   \includegraphics[width=3cm,clip=true,keepaspectratio=true]{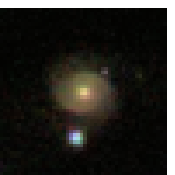}
   \includegraphics[width=3cm,clip=true,keepaspectratio=true]{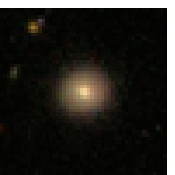}\\
   \includegraphics[width=3cm,clip=true,keepaspectratio=true]{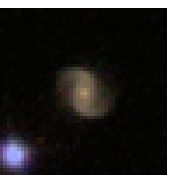}
   \includegraphics[width=3cm,clip=true,keepaspectratio=true]{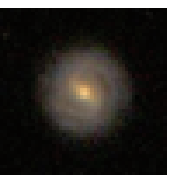}
   \includegraphics[width=3cm,clip=true,keepaspectratio=true]{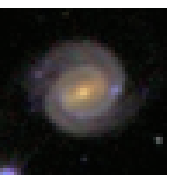}
   \includegraphics[width=3cm,clip=true,keepaspectratio=true]{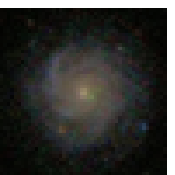}
   \includegraphics[width=3cm,clip=true,keepaspectratio=true]{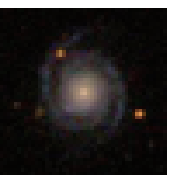}\\
   \includegraphics[width=3cm,clip=true,keepaspectratio=true]{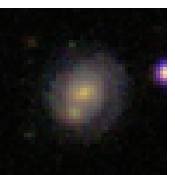}
   \includegraphics[width=3cm,clip=true,keepaspectratio=true]{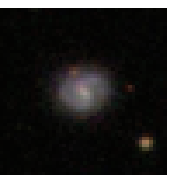}
   \includegraphics[width=3cm,clip=true,keepaspectratio=true]{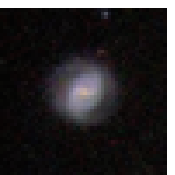}
   \includegraphics[width=3cm,clip=true,keepaspectratio=true]{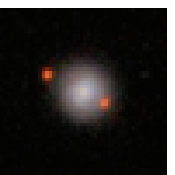}
   \includegraphics[width=3cm,clip=true,keepaspectratio=true]{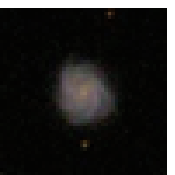}
   \caption{Examples of galaxies in our sample. The four rows show from top to bottom, respectively:
elliptical galaxies, galaxies with classical bulges, galaxies with pseudo-bulges, and bulgeless galaxies.
For the three last rows, the first three columns from left show barred galaxies. This grouping is based on the
results from our parametric image decompositions.}
   \label{fig:sample}
\end{figure*}

Figure \ref{fig:sample} shows examples of galaxies in our sample. Each row corresponds to a different galaxy
category, concerning the central component. Elliptical galaxies
are shown in the top row. The remaining rows show,
from top to bottom, respectively, galaxies with classical bulges, galaxies with pseudo-bulges, and bulgeless galaxies.
This grouping is based on the results from our parametric image decompositions below.

\section{Image Decomposition}
\label{sec:method}

The fitting of each galaxy image in our sample was done using {\sc budda} v2.1, which is able to fit up to four
different galactic components to a galaxy image, namely bulge, disc, bar, and a central source, usually adopted
to account for the emission from bright, type 1 AGN. Details on the code technical
aspects and capabilities can be found in \citet{deSGaddos04} and \citet{Gad08b}. In particular, the latter
discusses more specifically the treatment of the bar component. Here we briefly recall the functions used to
describe each component.

The model components are described by concentric, generalised ellipses \citep[see][]{AthMorWoz90}:

\begin{equation}
\left(\frac{|x|}{a}\right)^c+\left(\frac{|y|}{b}\right)^c=1,
\end{equation}

\noindent where $x$ and $y$ are the pixel coordinates of the ellipse points, $a$ and $b$ are the extent of its
semi-major and semi-minor axes, respectively, and $c$ is a shape parameter. The centre of the component, given
by the coordinates $x_0$ and $y_0$, is also fitted by the code, and is the same for all components fitted.
Position angles and ellipticities were fitted by the code for every component, each one having a single
position angle and ellipticity. When $c=2$ one has a simple, pure ellipse.
When $c<2$ the ellipse is discy, and when $c>2$ the ellipse is boxy. Bulges and discs were fitted using pure
ellipses, but $c$ was left free to fit bars, since bars are better described by boxy ellipses.

Each component also has to follow a surface brightness radial profile.
The disc profile is described by an exponential function \citep{Fre70},

\begin{equation}
\mu_d(r)=\mu_0 + 1.086r/h,
\end{equation}

\noindent where $r$ is the galactocentric radius, $\mu_0$ is the disc central surface brightness,
and $h$ is the disc characteristic scale-length.

The bulge profile is described by a S\'ersic function \citep[][see \citealt{CaoCapDOn93}]{Ser68},

\begin{equation}
\mu_b(r)=\mu_e+c_n\left[\left(\frac{r}{r_e}\right)^{1/n}-1\right],
\end{equation}

\noindent where $r_e$ is the effective radius of the bulge, i.e. the radius that contains half of its
light, $\mu_e$ is the bulge effective surface brightness, i.e. the surface brightness at $r_e$, $n$ is the
S\'ersic index, defining the shape of the profile, and $c_n=2.5(0.868n-0.142)$. When $n=4$ the profile
is a de Vaucouleurs function \citep{deV48}, while $n=1$ corresponds to an exponential bulge. Elliptical galaxies
were fitted with a single ``bulge'' component.

The bar profile is also described by a S\'ersic function. For the bar,

\begin{equation}
\mu_{\rm bar}(r)=\mu_{e,{\rm bar}}+c_{n,{\rm bar}}\left[\left(\frac{r}{r_{e,{\rm bar}}}\right)^{1/n_{{\rm bar}}}-1\right],
\end{equation}

\noindent where $c_{n,{\rm bar}}=2.5(0.868n_{{\rm bar}}-0.142)$, and the other parameters have similar definitions
as for the bulge. Another bar parameter fitted by the code is the length of the bar semi-major axis, $L_{{\rm bar}}$,
after which the bar profile is simply truncated and drops to zero.

Using the models obtained with {\sc budda}, we have calculated the absolute magnitude of each
different galaxy component in all galaxies in the sample in the $g$, $r$ and $i$ bands. We have thus
also calculated the integrated colours of each component separately.

To avoid unreliable results that very often arise from automated image decomposition, we applied {\sc budda}
to each galaxy individually. Erroneous fits are particularly likely to arise, if not followed closely, when
fitting barred galaxies. In such cases, the number of parameters fitted can be as high as 18, and the topology of
the distribution of $\chi^2$ for the many possible models is very complex.
We thus adopted a procedure to produce each fit, which is as follows.

First, we needed to produce the image to be fitted. To this end, a cutout was created from the original SDSS galaxy
image, with the galaxy in the centre, and containing enough sky background in order not to remove the galaxy outskirts.
Usually, this means that the image size on a side is about twice the apparent diameter of the galaxy.
Before fitting, each image was cleaned of spurious objects such as foreground stars.
The sky background and PSF FWHM values given by SDSS were checked and corrected when necessary, which rarely
occurred. When needed, the sky background was estimated by calculating the median pixel value of several areas free
from objects in the galaxy image. When the PSF FWHM given by SDSS was deemed erroneous, we have made new
estimates, by fitting a Moffat profile \citep{Mof69} to a number of
suitable stars near the galaxy, and choosing the median value found. The images were fitted without subtracting the
mean background value.
This is important, because a proper calculation of the $\chi^2$, and thus a reliable determination of the
uncertainties, is based on the statistics concerning all photons that reached the detector. Evidently, the value
of the background is given to the code, and the model components are determined taking it into
account.

Second, before starting the fit, it is necessary to provide the code with initial guesses for all parameters
to be fitted. The coordinates of the pixel with highest intensity were chosen as input to $x_0$ and $y_0$.
The geometrical parameters were roughly estimated by visual inspection of the image and
the corresponding isophotal contours. When deemed necessary, these parameters were also calculated using the intensity
moments in a few different circular apertures centred at the galaxy centre.
Finally, the initial guesses for the parameters of the surface brightness profiles were estimated through the
inspection of a pixel-by-pixel radial intensity profile of the galaxy. Tests with artificial images of unbarred
disc galaxies with high and low spatial resolution in \citet{deSGaddos04} showed that {\sc budda} is generally able
to reliably retrieve the correct structural parameters. In \citet{Gad08b}, further tests with similar
images showed that the uncertainties estimated by the code for each fitted parameter are realistic.
In Appendix \ref{sec:app}, we perform new tests, and check how the fits in the present study depend on the input
parameters, using SDSS $r$-band images of twenty barred galaxies in our sample. The choice for barred galaxies
assures that the results from these tests concern typically difficult fits. The results are encouraging.
They show that, in most cases, the code converges to similar results (generally consistent within the estimated
uncertainties) for input parameters varying up to a factor of four. Furthermore, wrong fits can in general be
recognised by their unusually high $\chi^2$ values.

A critical point at this stage is to decide on which components the model of the galaxy should include.
With this aim, the inspection of the galaxy image, isophotal contours and intensity profile was also very helpful.
A bulge component was included in the model if the intensity profile is clearly not a single
exponential function. In the absence of bars or
spiral arms, it might not be clear only by inspecting the image if a galaxy has or not a disc component.
The latter might become easier to identify when checking the isophotal contours for significant differences
in position angle and/or ellipticity between the inner and outer parts of the galaxy. In addition, an outer
exponential component in the intensity profile is naturally also an indication of the presence of a disc.
A disc was included in the model whenever one of such signatures was found.
To check for the presence of a bar, we first inspected the galaxy image. If in
doubt, we then looked for the typical bar signatures in the isophotal contours and the intensity profile, i.e.,
respectively, an elongated structure with constant position angle, and a flat ledge in the profile
\citep[see e.g.][]{GadAthCar07}. A bright AGN could be discerned if the profile has a cusp in the centre.
However, even bright AGN are smoothed out if the physical spatial resolution of the image is not high enough
\citep[see][]{Gad08b}. Accordingly, none of our images present such a feature, and thus no galaxy in our
sample required an AGN component in the fit.

Evidently, however, some cases are less clear, and one can not rule out the possibility of us fitting
a truly elliptical galaxy with a model containing bulge and disc, or conversely, fitting
an unbarred lenticular galaxy
with only a bulge model. In some of the latter cases, a first fit resulted in a bad match to the galaxy
image, usually a bulge with an obviously
too low S\'ersic index. In most of these cases, another fit, including a disc component, produced
a better match, revealing the true nature of the galaxy.
To check that our decision on whether or not adding a disc component in a given fit is reliable,
we have performed the following tests. We randomly selected 20 galaxies judged as ellipticals and
produced test fits with a disc component. Conversely, we randomly selected 20 galaxies judged as
unbarred lenticulars and produced test fits without a disc component. It turned out that, for
the ellipticals, only one test fit resulted in a lower $\chi^2$, and, for the lenticulars, two test
fits resulted in similar $\chi^2$ values. Nevertheless, none of such cases are significantly
improved fits, and most of these test fits resulted in unmistakably worse matches to the galaxy images.
Furthermore, the disc component found in the tested ellipticals never constributes with more than a
few per cent of the total galaxy luminosity.
It should also be mentioned that, due to the limited physical spatial resolution
of our images, we likely missed most bars shorter than $L_{\rm bar}\approx2-3$ kpc, typically seen in very late-type
spirals (later than Sc -- \citealt{ElmElm85}). The whole of these faint bars is typically within $2-4$ seeing
elements, and thus they do not imprint clear signatures in either the isophotal contours or the intensity profile.
Note that, in contrast, bulges with $r_e$ of the order of one seeing HWHM can still be identified in the
intensity profile, since they usually contain a much larger fraction of the galaxy light than these short and faint
bars.

This procedure was performed using several suitable tasks in {\sc iraf}.\footnote{{\sc iraf} is distributed by
the National Optical Astronomy Observatories, which are operated by the Association of Universities for Research
in Astronomy, Inc., under cooperative agreement with the National Science Foundation.}
The fits to each galaxy were first done in the $r$-band, until a satisfactory model is achieved.
Then the same input parameters which resulted in the final $r$-band fit were used to fit the $g$ and
$i$-band images, with
the luminosity parameters scaled accordingly. In addition, in some cases, geometrical parameters such as ellipticities
and position angles were kept fixed at the values found with the $r$-band image. A model was deemed incorrect if
any of the fitted parameters assumed a value which is clearly wrong, but this was uncommon.
A typical case concerns the bulge S\'ersic index.
If $n$ was found to be outside the range $0.8\leq n\leq6$, then a new set of input parameters was tried in a new fit.
Usually, the subsequent fits returned values for $n$ within this range, as well as smaller uncertainties in the fitted
parameters, and a lower $\chi^2$. In these cases, we sticked to the new fit. In some other cases, however, the
resulting model did not change after changing the initial parameters. When this happened, we considered the model
appropriate. Only occasionally a more detailed comparison between galaxy image and model was done, e.g. by
comparing both intensity profiles.

\begin{figure*}
   \centering
\vskip -0.6cm
   \includegraphics[width=4cm,clip=true,keepaspectratio=true]{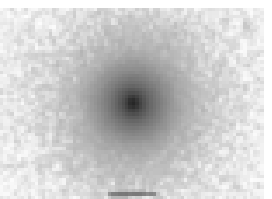}
   \includegraphics[width=4cm,clip=true,keepaspectratio=true]{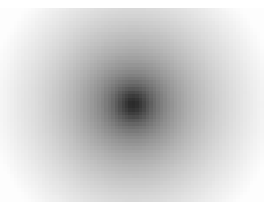}
   \includegraphics[width=4cm,clip=true,keepaspectratio=true]{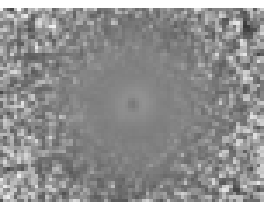}
   \includegraphics[width=4cm,clip=true,keepaspectratio=true]{0382-51816-363_prof.eps}
   \includegraphics[width=4cm,clip=true,keepaspectratio=true]{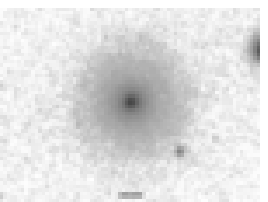}
   \includegraphics[width=4cm,clip=true,keepaspectratio=true]{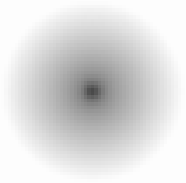}
   \includegraphics[width=4cm,clip=true,keepaspectratio=true]{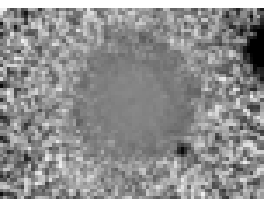}
   \includegraphics[width=4cm,clip=true,keepaspectratio=true]{0875-52354-113_prof.eps}
   \includegraphics[width=4cm,clip=true,keepaspectratio=true]{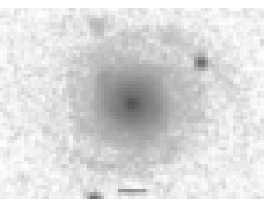}
   \includegraphics[width=4cm,clip=true,keepaspectratio=true]{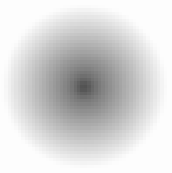}
   \includegraphics[width=4cm,clip=true,keepaspectratio=true]{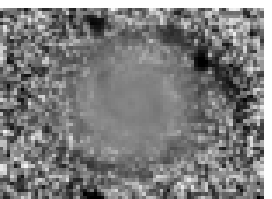}
   \includegraphics[width=4cm,clip=true,keepaspectratio=true]{0467-51901-283_prof.eps}
   \includegraphics[width=4cm,clip=true,keepaspectratio=true]{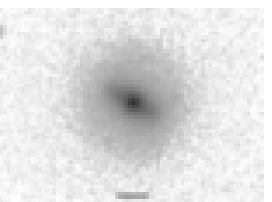}
   \includegraphics[width=4cm,clip=true,keepaspectratio=true]{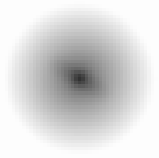}
   \includegraphics[width=4cm,clip=true,keepaspectratio=true]{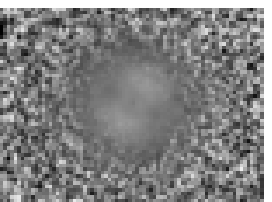}
   \includegraphics[width=4cm,clip=true,keepaspectratio=true]{0789-52342-340_prof.eps}
   \includegraphics[width=4cm,clip=true,keepaspectratio=true]{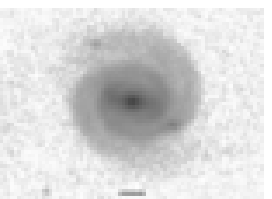}
   \includegraphics[width=4cm,clip=true,keepaspectratio=true]{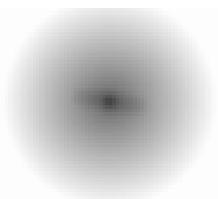}
   \includegraphics[width=4cm,clip=true,keepaspectratio=true]{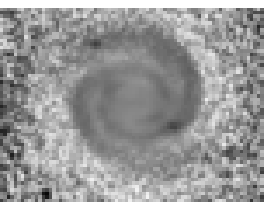}
   \includegraphics[width=4cm,clip=true,keepaspectratio=true]{0543-52017-026_prof.eps}
   \includegraphics[width=4cm,clip=true,keepaspectratio=true]{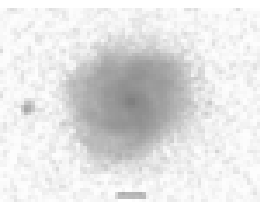}
   \includegraphics[width=4cm,clip=true,keepaspectratio=true]{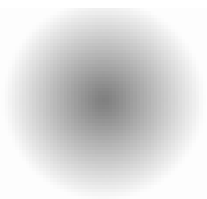}
   \includegraphics[width=4cm,clip=true,keepaspectratio=true]{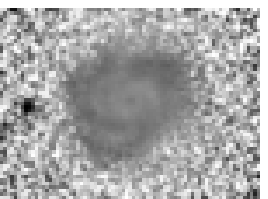}
   \includegraphics[width=4cm,clip=true,keepaspectratio=true]{0575-52319-274_prof.eps}
\vskip -0.15cm
   \caption{Examples of fits obtained. Each row refers to a single galaxy, showing, from left to right:
the original $i$-band galaxy image, the model image obtained with {\sc budda}, the residual image, and
surface brightness profiles. Horizontal lines in the galaxy images mark a length of 5 kpc at the
galaxy distance. The first two images have intensities transformed to magnitudes per square arcsecond, and
are displayed with the same levels in all rows. The residual images are obtained subtracting the model image
from the original image, and are displayed with a much narrower range in intensity, which is also the same in all
rows, in order to enhance residual sub-structures. In the residual images, darker pixels indicate where
the galaxy is brighter than the model, whereas whiter pixels indicate where the model is brighter than the galaxy.
The surface brightness profiles were obtained from cuts along the galaxy major axis, or along the bar major axis
when there is a bar. The dashed line corresponds to the original galaxy image, whereas the black solid line
corresponds to the model. When there is more than a single component, red, blue and green lines refer,
respectively, to bulge, disc and bar. The dotted line in the bottom panel shows the residuals (galaxy $-$ model).
These results correspond to, from top to bottom, respectively: an elliptical galaxy, a galaxy with a classical bulge, a
galaxy with a pseudo-bulge, a barred galaxy with a classical bulge, a barred galaxy with a pseudo-bulge,
and a bulgeless galaxy.}
   \label{fig:decomps}
\end{figure*}

Figure \ref{fig:decomps} shows examples of the fits obtained for six galaxies in the sample. These include
an elliptical galaxy, two galaxies with a classical bulge (one barred galaxy and one unbarred),
two galaxies with a pseudo-bulge (again barred and unbarred), and a bulgeless galaxy.
The distinction between classical bulges and pseudo-bulges is explained in details in Sect. \ref{sec:pseudo}.
One sees that the fits are generally quite good, even though the physical spatial resolution of the images is restricted,
and despite the fact that, given the size of our sample, we could not make a thourough comparison between galaxy and
model before accepting the fit to every single galaxy in the sample. This would require an unreasonable amount of time
to complete the study for the whole sample. The discrepancies between galaxy and model are usually below
0.3 mag arcsec$^{-2}$ and are mainly caused by sub-structures such as spiral arms, rings,
lenses and bar ansae. It is clear that,
if we can not ascertain that the fits are excellent for a single specific galaxy, the results from the fits are, on average,
trustworthy, and thus suitable for statistical studies.

\section{Results}
\label{sec:results}

\subsection{A comparison between different bulge-to-total indicators}

The morphological (Hubble) type of a galaxy is related to several physical parameters that are an indication
of the galaxy formation and evolution processes \citep[e.g.][]{RobHay94}.
One of the basic features on which the morphological classification of a galaxy hinges on is the relative
importance of the bulge component to the overall light distribution of the galaxy \citep{Hub26}.
Thus, it is of the utmost importance to be able to measure how relevant is the bulge component within a
given galaxy in an objective and quantitative fashion, even if there is some scatter in the relation between
measured bulge prominence and visually determined Hubble type \citep[e.g.][]{LauSalBut07,GraWor08}. Because
it is difficult to objectively define the latter, the former is advantageous.

\begin{figure}
   \centering
   \includegraphics[width=7cm,clip=true,keepaspectratio=true]{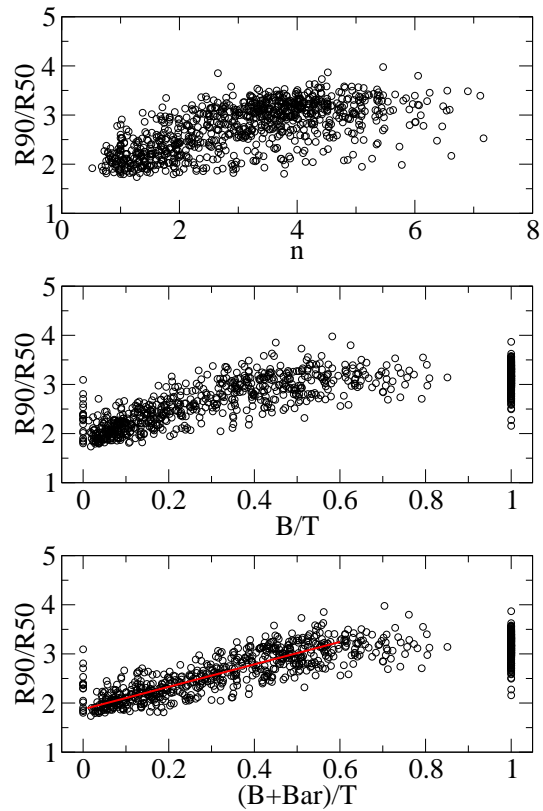}
   \caption{Concentration index ($R90/R50$) plotted against bulge S\'ersic index $n$ (top),
bulge-to-total ratio $B/T$ (middle) and the
added contributions of bulge and bar to the total galaxy luminosity $(B+Bar)/T$ (bottom). The correlation between concentration
and $n$ is remarkably poorer than that with $B/T$. Interestingly, the latter correlation is improved when the bar
contribution is taken into account. Elliptical galaxies and bulgeless galaxies can be readily identified in the middle
panel at $B/T=1$ and $B/T=0$, respectively. The red line in the bottom panel is a linear fit to the data points with
$0<(B+Bar)/T\leq0.6$. For galaxies with $B/T$, or $(B+Bar)/T$, greater than $\approx0.6$, $R90/R50$ is roughly constant at
$\approx3$.}
   \label{fig:conc}
\end{figure}

\begin{figure}
   \centering
   \includegraphics[width=8cm,clip=true,keepaspectratio=true]{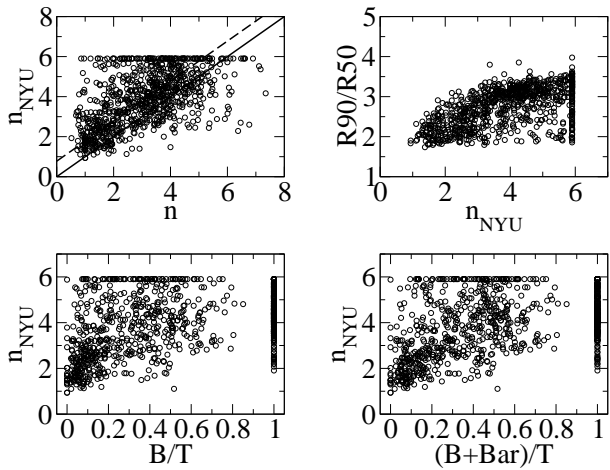}
   \caption{Top left: {\em galaxy} S\'ersic index from \citet{BlaEisHog05} plotted against
our measurements of {\em bulge}
S\'ersic index; top right: concentration index plotted against galaxy S\'ersic index; bottom left:
galaxy S\'ersic index plotted against $B/T$; bottom right: galaxy S\'ersic index plotted against $(B+Bar)/T$.
The solid line in the top left panel depicts a perfect correspondence. The dashed line is a linear fit to all data points,
fixing the slope of the fit to 1. Although there is a very large spread, it is clear that galaxy S\'ersic index is
systematically higher than bulge S\'ersic index by, on average, $\approx0.8$.}
   \label{fig:conc_nyu}
\end{figure}

In this work, since we have performed a detailed decomposition of each galaxy image into its main components,
bulge prominence is readily accessible from the bulge-to-total luminosity ratio $B/T$, calculated from
the models obtained with {\sc budda}. However, the employment of such techniques to very large samples
of galaxies presents serious challenges, as automated procedures are prone to large uncertainties. Thus,
other parameters, which are not so difficult to determine, have been suggested as a measure
of bulge prominence. One of such parameters
is the Petrosian concentration index, defined as the ratio $R90/R50$, where $R90$ and $R50$ are, respectively,
the radii enclosing 90 and 50 per cent of the galaxy luminosity \citep[see][and references therein]{KauHecWhi03}.
Another possibility is to fit the entire galaxy as a single component, with a S\'ersic function, and then
use the S\'ersic index so obtained \citep[e.g.][]{BlaSchStr05,BlaEisHog05,HauMcIBar07}.

Since we have measurements of $B/T$ and bulge S\'ersic index for all galaxies in our sample,
we can test how well concentration index
and galaxy S\'ersic index relate to $B/T$, and how all these connected parameters
relate to each other. This is done in Figs. \ref{fig:conc} and
\ref{fig:conc_nyu}. Concentration index is taken from the SDSS database and galaxy S\'ersic index from
\citet{BlaEisHog05}. The latter are updated values after the finding of a systematic bias in the values in
\citet{BlaHogBah03}. Figure \ref{fig:conc} shows that concentration correlates poorly with bulge S\'ersic
index. Although one sees a tendency in the sense that bulges with higher values for $n$ reside in galaxies
with higher $R90/R50$, the spread is very large. This might be partly due to the fact that $n$
is related only to the bulge component, whereas $R90/R50$
depends on the overall structure of the whole galaxy.
Another issue is that $n$ can have high uncertainties compared to its dynamical range. Figure \ref{fig:conc}
also shows that $R90/R50$ is well correlated with $B/T$. Interestingly,
the correlation gets noticeably better when the bar contribution to the total galaxy luminosity is taken into
account. The correlation coefficient for $R90/R50\times B/T$ is 0.82, while that for $R90/R50\times (B+Bar)/T$
is 0.86. Although bars are more extended structures than bulges, they are usually significantly less extended than
discs, which explains the better correlation of $R90/R50$ with $(B+Bar)/T$. This means that there is an
additional uncertainty when using $R90/R50$ as a proxy for $B/T$, as one galaxy with small bulge but
massive bar can have a higher value of $R90/R50$ than an unbarred, but otherwise similar, galaxy.
It is worth noticing that concentration rises linearly with $(B+Bar)/T$ up to $(B+Bar)/T\approx0.6$, after which
it stays roughly constant at $\approx3$, a value of concentration similar to that of massive elliptical galaxies.
A linear fit to the data points with $0<(B+Bar)/T\leq0.6$ gives

\begin{equation}
R90/R50=1.93(\pm0.02)+2.02(\pm0.05)\times (B+Bar)/T \textrm{.}
\end{equation}

Figure \ref{fig:conc_nyu} shows how galaxy S\'ersic index relates to bulge S\'ersic index, concentration,
$B/T$ and $(B+Bar)/T$. All plots show a very large spread. In particular, galaxy S\'ersic index is
systematically higher than bulge S\'ersic index by, on average, $\approx0.8$. Moreover, galaxy
S\'ersic index is a poor proxy for $B/T$. From the previous figure one sees that concentration
index is much more reliable in this respect. The correlation coefficients for galaxy S\'ersic index
as a function of $B/T$ and $(B+Bar)/T$ are, respectively, 0.43 and 0.45.

Nevertheless, it should be noted that both the galaxy and bulge S\'ersic indices in Figs. \ref{fig:conc}
and \ref{fig:conc_nyu} are corrected for seeing effects, whereas concentration index is not. It does
seem, however, that seeing effects on concentration index are small
at low redshifts \citep{BlaHogBah03}. For a sample with a
narrow range in redshift, as ours, such effects seem to be practically negligible, since the systematic variation
of the average physical size of the PSF is relatively small. We verified this by plotting $R90/R50$ against
redshift and finding no correlation whatsoever. Seeing effects would only introduce an uncertainty at each
measurement, which is small compared to the spread in concentration seen at each $B/T$, or $(B+Bar)/T$,
in Fig. \ref{fig:conc}.

\subsection{Identifying pseudo-bulges}
\label{sec:pseudo}

\begin{figure}
   \centering
   \includegraphics[width=7cm,clip=true,keepaspectratio=true]{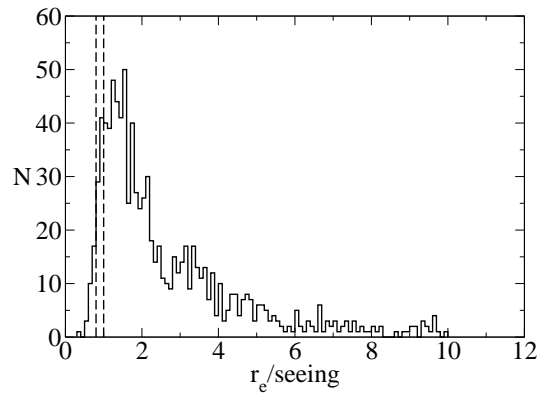}
   \caption{Distribution of the ratio between bulge effective radius and PSF HWHM for all galaxies in the sample
(excluding bulgeless galaxies). The two vertical dashed lines mark the positions where this ratio is 0.8 and 1.
Only 3 per cent (10 per cent) of the bulges have effective radii below 0.8 (1) times the PSF HWHM.}
   \label{fig:reseeing}
\end{figure}

\begin{figure*}
   \centering
   \includegraphics[width=10cm,clip=true,keepaspectratio=true]{car99_meanmue_new.eps}
   \includegraphics[width=7cm,clip=true,keepaspectratio=true]{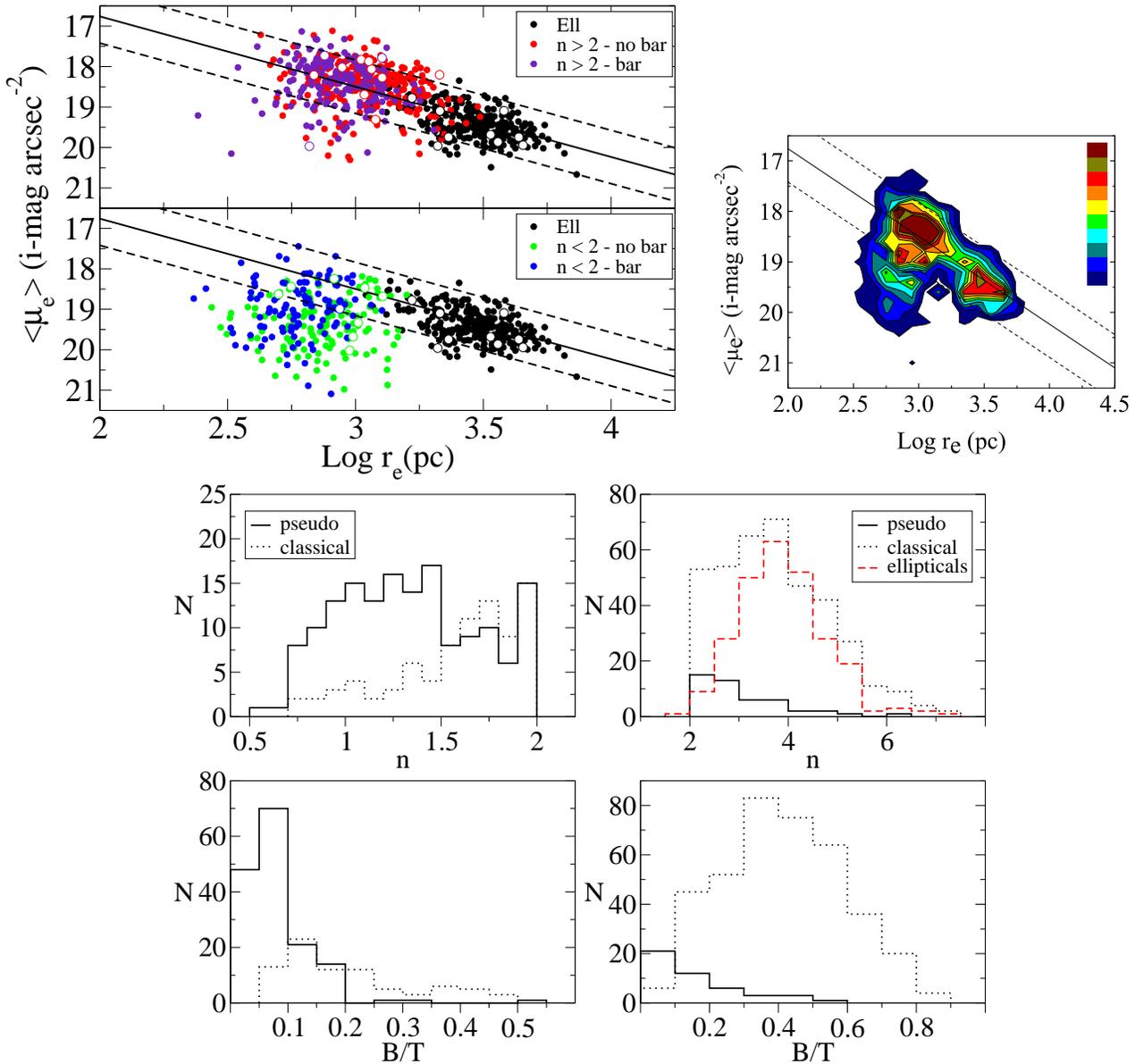}\\
   \includegraphics[width=6cm,clip=true,keepaspectratio=true]{car99_hists_pse_new.eps}
   \includegraphics[width=6cm,clip=true,keepaspectratio=true]{car99_hists_cla_new.eps}
   \caption{Graphical representation of our criterion to distinguish classical and pseudo-bulges. The two
contiguous panels at the top show the mean effective surface brightness {\em within} the effective
radius plotted against the logarithm
of the effective radius. The two panels display elliptical galaxies and, separately, bulges with S\'ersic
index above and below 2. Barred and unbarred galaxies are also indicated. The solid line is a fit to the
elliptical galaxies and the two dashed lines point out its $\pm3\sigma$ boundaries. Bulges that lie below
the lower dashed line are classified as pseudo-bulges, {\em independently} of their S\'ersic index.
Bigger, white-filled circles represent systems with $0.5<b/a<0.7$, with similar colour coding.
They do not show any clear distinct patterns. The separate panel at the top shows a density
plot with all data points taken altogether. It shows that ellipticals, classical and pseudo-bulges
indeed appear as independent groups in the Kormendy relation. The four histograms at the bottom show the distributions
of S\'ersic index $n$ and bulge-to-total luminosity
ratio $B/T$ for classical bulges and pseudo-bulges with $n<2$ (left) and $n\geq2$ (right). While the threshold
in S\'ersic index can be considered as an approximation to identify pseudo-bulges, it is clear that
it can generate many misclassifications. Our criterion can recognise more precisely pseudo-bulges, as those which are
structurally different from classical bulges and ellipticals (also shown in the top right histogram).}
   \label{fig:car99}
\end{figure*}

As mentioned in the Introduction, studies on the properties of pseudo-bulges are often characterised by having
a physical spatial resolution higher than that of the SDSS images used here. It is thus legitimate to be concerned
on how well can we in fact distinguish pseudo-bulges from classical bulges. One key aspect of this assessment
is the comparison between the effective radius of the bulge $r_e$ and the PSF HWHM of the corresponding image.
\citet{Gad08b} showed that the structural properties of bulges can be reliably retrieved provided that $r_e$ is larger
than $\sim80$ per cent of the PSF HWHM. This was gauged by fitting images of very nearby galaxies, at relatively
high physical spatial resolution, and through the fitting of the same galaxies using artificially redshifted
images. Figure \ref{fig:reseeing} shows that only three per cent of our bulges
do not satisfy this criterion. If one wants to be more stringent and extend this limit to 100 per cent of
the seeing radius, one still sees that only 10 per cent of our bulges are below the threshold. We have not
excluded these bulges from this study but have verified that their inclusion or removal does not result in any
significant change in the results presented here. This gives us confidence that the structural properties
we have obtained are generally reliable even for the less conspicuous bulges.
Note that our decompositions are seeing-corrected.

The top two panels in Fig. \ref{fig:car99} show the mean effective surface brightness within the effective
radius $\left<\mu_e\right>$\footnote{Note that $\left<\mu_e\right>$ is not equal to $\mu_e$, which is the
effective surface brightness {\em at} the effective radius. In plots such as Fig. \ref{fig:car99} and the fundamental
plane (discussed further below),
$\left<\mu_e\right>$ must be used, since it takes into account differences in the light
(mass) distribution within different systems.}
plotted against effective radius for the bulges and elliptical galaxies in our sample. Massive elliptical galaxies
follow a well defined relation between $\left<\mu_e\right>$ and $r_e$, which is given by the solid line, constructed from
a fit to the data. The two dashed lines are the fit's $\pm3\sigma$ boundaries (over the zero-point and keeping
the slope fixed). This relation is known as the Kormendy relation \citep{Kor77} and is in fact one of the possible
projections of the fundamental plane (we explore the fundamental plane relations in more detail in Sect.
\ref{sec:FP}). The slope of the Kormendy relation we find here is similar (consistent within the uncertainties)
to that found by \citet{PieGavFra02} using $H$-band images. Many bulges also follow this
relation. If bulges are separated according to their S\'ersic index $n$, one sees that most of the bulges with $n\geq2$
follow the ellipticals' relation, whereas most of those with $n<2$ do not, lying below the relation.
This is expected, since exponential
bulges are often pseudo-bulges, which are believed to be structurally different from classical bulges and ellipticals
\citep[see e.g.][and references therein]{Car99,KorKen04}. However, many of the bulges with $n<2$ do seem to follow
the same relation of ellipticals. Conversely, several bulges with $n\geq2$ fall below this relation. It is thus
clear that although the S\'ersic index threshold can be considered as an approximation to identify
pseudo-bulges \citep[see e.g.][]{FisDro08},
it can introduce many misclassifications. Since pseudo-bulges are expected to be structurally different
from classical bulges, and to lie below the $\left<\mu_e\right>-r_e$
relation for ellipticals, we can use the lower dashed line
in Fig. \ref{fig:car99} to be the dividing line between classical bulges and pseudo-bulges. This means that
pseudo-bulges satisfy the following inequality:

\begin{equation}
\left<\mu_e\right> > 13.95+1.74\times\log r_e\textrm{,}
\end{equation}

\noindent where $\left<\mu_e\right>$ and $r_e$ are measured in the SDSS $i$-band, and $r_e$ is in parsecs.
In fact, in this diagram, discs also occupy the region below the relation defined by ellipticals \citep{PieGavFra02},
which is consistent with the expectation that pseudo-bulges are built from disc instabilities.
From Fig. \ref{fig:car99} one also sees that barred and unbarred galaxies behave in a similar fashion in what
concerns their positions in the $\left<\mu_e\right>-r_e$ relation.
The histograms in Fig. \ref{fig:car99} show the distributions of $n$ and $B/T$ for classical bulges
and pseudo-bulges with $n<2$ and $n\geq2$ in separate panels. They show that our criterion to
separate these two bulge categories can identify bulges with $n<2$, but with high values of $B/T$, as
classical bulges. These bulges have often values of $n$ close to 2. Conversely, we are also able
to identify pseudo-bulges with $n\geq2$. These bulges show often low values of $B/T$ and also values for $n$
close to 2. These results indicate that some bulges can be erroneously classified as classical or pseudo, if one
uses the S\'ersic index for such classification, just for being by chance (due to measurement uncertainties)
in the wrong side of the S\'ersic index threshold. One sees that, as expected, pseudo-bulges
show a distribution of $n$ unlike that of classical bulges and ellipticals. In addition, the distribution
of $n$ for classical bulges is wide, and ranges from values typical of ellipticals to values in the gap between
pseudo-bulges and ellipticals. Furthermore, one also sees that classical bulges
have a peak in $B/T$ around 0.4, while the corresponding value for pseudo-bulges is about 0.1.

Although it is clear that bulges with low and high S\'ersic index occupy different loci in the $\mu_e-r_e$
projection, Fig. \ref{fig:car99} suggests that there is a more significant difference between all bulges
and ellipticals than between the different bulges. To check that, we have performed 2D Kolmogorov-Smirnov
tests \citep{FasFra87,PreTeuVet92}. The tests show that, in fact, {\em in this projection}, ellipticals
differ more from bulges (taken altogether) than bulges with high S\'ersic index differ from bulges with low
S\'ersic index. In this context, we have checked, using the tests described in Sect. \ref{sec:method},
that if a lenticular galaxy is mistaken by an elliptical, the shift in $r_e$ could be at most only 0.04 dex.
Thus, although such an effect, if present, would indeed bring ellipticals closer to bulges in the
Kormendy relation, its amplitude is too small. In light of these results, one could ask whether
a separation between classical and pseudo-bulges in the Kormendy relation is rather artificial.
The density plot in Fig. \ref{fig:car99} suggests that it is not: pseudo-bulges appear as an
independent group of points, as do classical bulges and ellipticals.

\begin{figure}
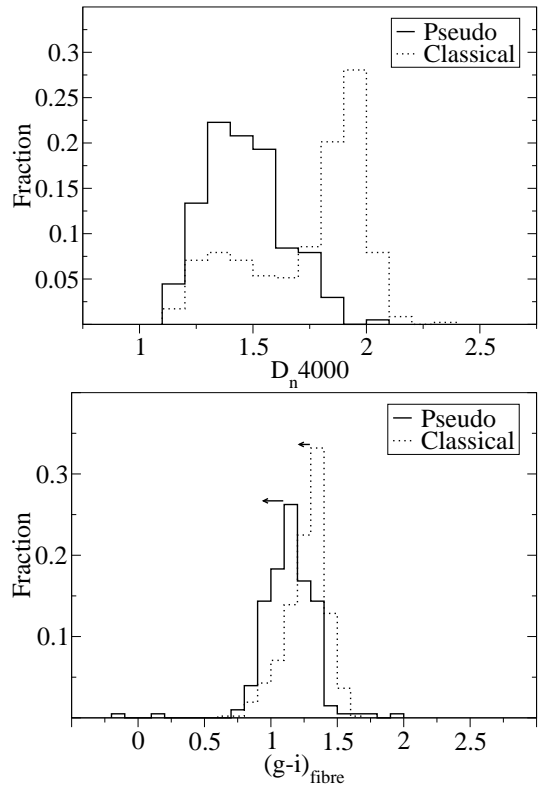

   \centering
   \includegraphics[width=7cm,clip=true,keepaspectratio=true]{d4.eps}\\
   \includegraphics[width=7cm,clip=true,keepaspectratio=true]{fibcol.eps}
   \caption{Distributions of D$_n$(4000) (top) and fibre $g-i$ colour (bottom) for pseudo-bulges and classical bulges.
The arrows in the bottom panel indicate an upper limit on how dust reddening could affect the colours of each bulge
category, on average.}
   \label{fig:d4col}
\end{figure}

Previous studies also indicate that pseudo-bulges show star-forming activity, as opposed to classical bulges,
which are constituted mainly by old stellar populations
\citep[e.g.][]{Gaddos01,Fis06}. We can check this difference within our sample
using both the D$_n$(4000) index provided in \citet{KauHecWhi03} and the SDSS fibre colour (i.e. the colour
measured within the fibre aperture). The D$_n$(4000) index is based on the 4000\AA\, discontinuity seen in optical
spectra of galaxies, and it is not sensitive to dust attenuation effects. Young stellar
populations have low values of D$_n$(4000), as compared to old stellar populations. The fibre colours
are integrated over the central region of the galaxy and can suffer from dust reddening.
In Fig. \ref{fig:d4col}, we show the distributions of D$_n$(4000) and fibre $g-i$ colour
for pseudo-bulges and classical bulges. The result is striking. Our pseudo-bulges not only do indeed show values of
D$_n$(4000) lower than those for classical bulges, indicating a higher level of star formation activity, but
their distributions are also significantly apart (except for a tail of classical bulges showing
star-forming activity). In fact, the peaks of the D$_n$(4000) distributions for pseudo-bulges and classical
bulges coincide with the peaks in the bimodal distribution of D$_n$(4000) seen in both our parent and final samples
(see Fig. \ref{fig:scomp}). Such bimodality was also found by \citet{AllDriGra06}, who concluded that
it is produced by bulge-dominated galaxies on the one hand, and disc-dominated galaxies on the other hand. More
recently, \citet{DroFis07} suggested that such bimodality is a result of the two different bulge families:
pseudo-bulges are hosted by galaxies with high overall star formation activity, whereas classical bulges
reside in galaxies that underwent violent relaxation processes in the past, such as mergers, and thus
have formed most of their stars in these past events. Both suggestions are not mutually exclusive, since
galaxies with pseudo-bulges are often disc-dominated, as opposed to bulge-dominated galaxies, which often
host classical bulges. In fact, our results indicate that the most disc-dominated galaxies almost always
host pseudo-bulges, which are almost always under current intense star-forming activity,
whereas classical bulges are mostly quiescent and reside mainly in more bulge-dominated galaxies.
Interestingly enough, star-forming classical bulges form a group of points of their own in the density
plot of Fig. \ref{fig:car99}, between the lower dashed line and the solid line.
We will come back to these points further below.
We have verified that if one separates bulges using their S\'ersic index with a threshold at $n=2$ the
resulting D$_n$(4000) distributions are similar, and one also sees a tail of star-forming bulges with $n\geq2$.
In this case, however, the peaks of such distributions
are slightly closer, and the distribution for bulges with $n<2$ is broader than that of our pseudo-bulges.
This gives support to our criterion to identify pseudo-bulges.

The colour distributions are also separated, though in a less dramatic way. Pseudo-bulges show
bluer colours than classical bulges, on average, but the difference in the mean colour values amounts
only to 0.2 mag. Evidently, dust reddening does play a role in making these distributions more similar, as compared
to those of D$_n$(4000). Some dust attenuation is expected even in face-on galaxies \citep[e.g.][]{DriPopTuf08}.
The fact that the colour distribution of pseudo-bulges is wider than that of
classical bulges is also a result of dust attenuation. On the one hand, there are those pseudo-bulges with not much dust
attenuation, whose intense star formation shows up as the resulting blue colours (and the blue tail in the distribution).
And, on the other hand, there are those pseudo-bulges with dust-obscured star formation that produce the red colours
seen in the red tail of the distribution. One can have an estimate of the dust reddening in our bulges,
using estimates of the total galaxy dust attenuation in \citet{KauHecWhi03} as an upper limit (i.e. considering that
all dust attenuation in the galaxy strikes the bulge light). We calculated the average dust
attenuation in the $i$-band using the results for the $z$-band in \citet{KauHecWhi03}, and assuming that dust
attenuation goes as $\lambda^{-0.7}$ (where $\lambda$ is the observed wavelength).
The average $i$-band attenuation we thus find for our elliptical galaxies and galaxies with
classical and pseudo-bulges is, respectively, 0.04, 0.16 and 0.48 mag. This corresponds to $g-i$ colour
excesses of, respectively, 0.02, 0.08 and 0.15 mag.
The colour excess for classical and pseudo-bulges are indicated by the arrows in the lower panel of Fig. \ref{fig:d4col}.
Interestingly, the difference in the position of the peaks of the D$_n$(4000) distributions amounts to
0.60, while the corresponding difference for fibre $g-i$ colour is, as mentioned above, 0.20 mag
(without any dust correction) and 0.27 (with the upper limit dust correction).

Since the D$_n$(4000) values and fibre colours are taken from SDSS spectra, which are themselves
taken through a {\em fixed} fibre aperture of 3 arcsec (thus 1.5 arcsec in radius),
and since pseudo-bulges are often smaller than classical bulges, one might ask whether
the striking difference in D$_n$(4000) between pseudo-bulges and classical bulges in our sample
is only due to aperture
effects. One could argue that, due to the smaller extent of pseudo-bulges within the fibre aperture, the
D$_n$(4000) values (and similarly the fibre colours)
for pseudo-bulges are contaminated with star formation in the disc. In fact, the average effective
radius of pseudo-bulges and classical bulges is, respectively, 0.82 arcsec and 1.14 arcsec. However, one must bear
in mind that bulge light dominates over disc light through $\approx$ 2
times the bulge effective radius from the galaxy centre. This is borne out in our surface brightness
profiles, and has also been found by others \citep[see e.g.][]{MorGioHay98,MorPomPiz08}.
Nevertheless, to be reassured that the difference seen in D$_n$(4000) is not just a spurious effect, we have
verified that the difference in the mean values of D$_n$(4000) for pseudo-bulges and classical bulges is similar
when all bulges are separated in bins of similar effective radius. In particular, we have verified that all results
from Fig. \ref{fig:d4col} hold if one considers only bulges with $r_e>1$ arcsec, or $r_e>1.5$ arcsec,
where little disc contamination is expected. Furthermore, we have also verified that the SDSS fibre colours
are in fact much more similar to the integrated bulge colours,
as calculated with the determined {\sc budda} models for the $g$ and $i$-band images,
than the corresponding disc colours. We thus conclude that one can not blame disc contamination
within the SDSS spectra to explain the higher star formation activity and bluer colours seen in our pseudo-bulges.

\begin{figure}
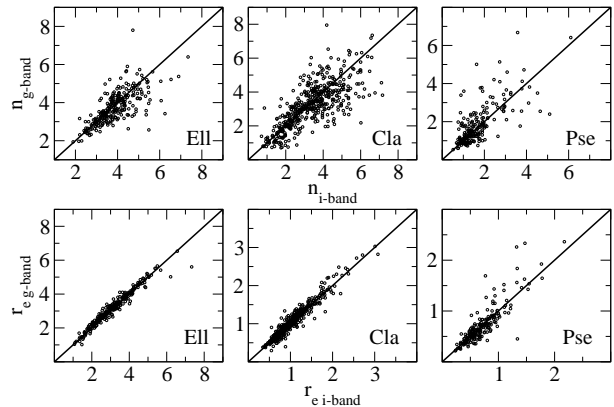

   \centering
   \includegraphics[width=8cm,clip=true,keepaspectratio=true]{n_all.eps}\\
   \includegraphics[width=8cm,clip=true,keepaspectratio=true]{re_all.eps}\\
   \caption{Comparison between structural parameters of elliptical galaxies and bulges obtained in the $g$ and
$i$-bands. Top: S\'ersic index; bottom: effective radius. Solid lines depict a one-to-one correspondence.
Bulges are separated according to our classification between classical and pseudo-bulges. The absence of systematic
effects indicates that dust does not play a significant role in the overall results of the decompositions.}
   \label{fig:dust}
\end{figure}

Our sample was designed in a way to minimise dust effects. In fact, most of our galaxies have dust attenuation
measurements in the low end of the distribution of dust attenuation for the whole sample of more than
$10^5$ galaxies in \citet{KauHecWhi03}. Nevertheless, it is clear that the high star formation
activity seen in pseudo-bulges is accompanied to some extent with dust obscuration. Could this somehow
affect the structural parameters obtained? To answer this question we plot in Fig. \ref{fig:dust}
bulge structural parameters obtained with the $g$-band images against the same parameters obtained
with the $i$-band images. Dust has more pronounced effects in bluer bands and thus, if there is some
systematic effects caused by dust in our decompositions, these should show up in these plots, in particular
for pseudo-bulges. For instance, dust attenuation could in principle result in systematically lower values for $n$,
as well as smaller values for $r_e$, in the $g$-band, as compared to those in the $i$-band. Figure \ref{fig:dust}
shows no substantial systematic effects. We thus conclude that our decompositions are not significantly disturbed
by dust. This is consistent with the findings in \citet{MorGioHay98}, who found that bulge structural
parameters do not change significantly with galaxy inclination.

\subsection{Scaling relations}
\label{sec:scale}

\begin{figure}
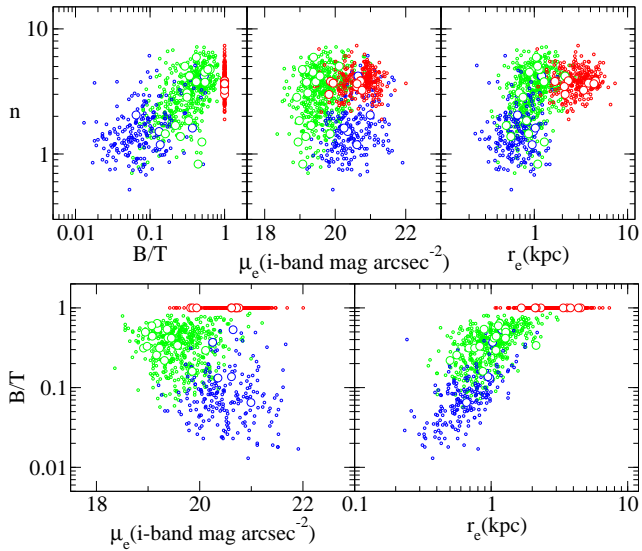

   \centering
   \includegraphics[keepaspectratio=true,width=8.4cm,clip=true]{nbtmur.eps}\\
   \includegraphics[keepaspectratio=true,width=8.4cm,clip=true]{btmur.eps}
   \caption{Bulge S\'ersic index plotted against bulge-to-total luminosity ratio (top left), bulge
effective surface brightness (top centre) and bulge effective radius (top right). And bulge-to-total
luminosity ratio plotted against bulge effective surface brightness (bottom left) and effective radius
(bottom right). Blue, green and red circles represent, respectively, pseudo-bulges, classical bulges and
ellipticals, with $b/a\geq0.9$. Bigger, white-filled circles with similar colour coding represent
systems with $0.5<b/a<0.7$. They follow similar patterns.}
   \label{fig:nbtmur}
\end{figure}

\begin{figure}
   \centering
   \includegraphics[keepaspectratio=true,width=8.4cm,clip=true]{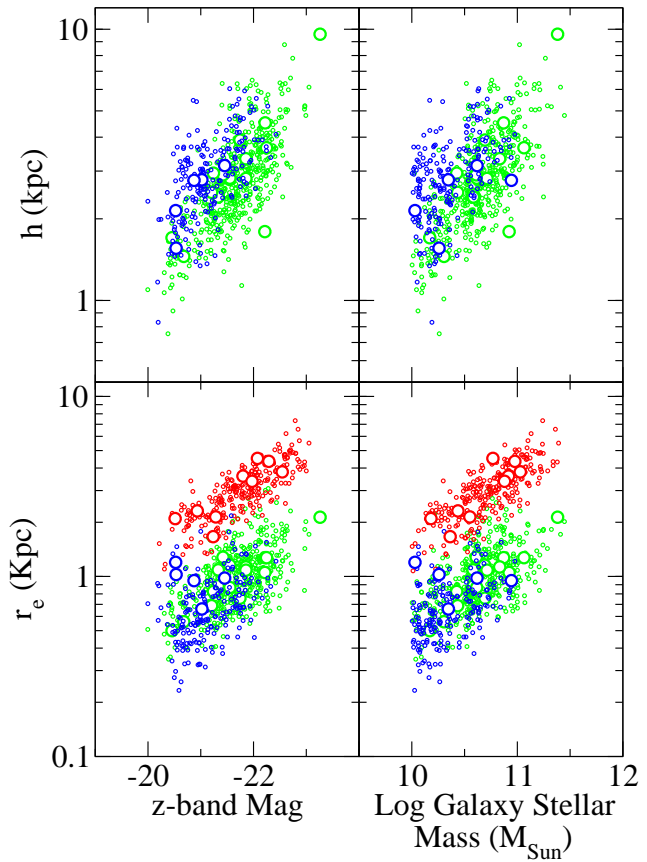}
   \caption{Disc scale-length and bulge effective radius plotted against galaxy $z$-band absolute magnitude
(k-corrected to $z=0.1$) and dust-corrected stellar mass, from \citet{KauHecWhi03}.
Blue, green and red circles represent, respectively, galaxies with pseudo-bulges, galaxies with classical
bulges and ellipticals, with $b/a\geq0.9$. Bigger, white-filled circles with similar colour coding represent
systems with $0.5<b/a<0.7$. They follow similar patterns.}
   \label{fig:hregalmagm}
\end{figure}

\begin{figure}
   \centering
   \includegraphics[keepaspectratio=true,width=8.4cm,clip=true]{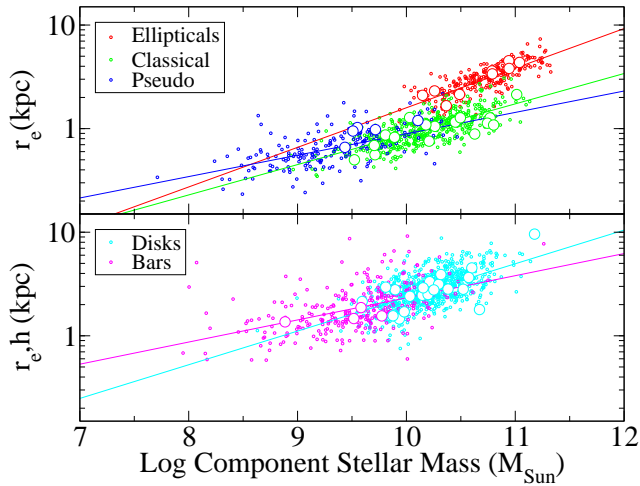}
   \caption{Scale-lengths of elliptical galaxies, classical bulges, pseudo-bulges, discs and bars,
plotted against the corresponding component stellar mass, as indicated. Solid lines are fits to the data.
Bigger, white-filled circles represent systems with $0.5<b/a<0.7$, with similar colour coding. They show
a similar behaviour.}
   \label{fig:scale}
\end{figure}

Figure \ref{fig:nbtmur} shows the relations between bulge S\'ersic index and bulge-to-total
luminosity ratio, effective surface brightness and effective radius for pseudo-bulges, classical
bulges and ellipticals. The relations between bulge-to-total luminosity ratio and effective surface
brightness and effective radius are also shown.
It is striking how these three stellar systems populate well-defined and
separate regions in some of these diagrams. Systems with larger $n$ tend to be more conspicuous and extended.
One sees a correlation between $n$ and $B/T$ for classical bulges but not as much for pseudo-bulges,
corroborating the results from \citet{FisDro08}, who arrived at a similar conclusion using
a smaller sample. It is worth noticing that $n$ rises with $r_e$ for bulges but it is rather constant for
ellipticals. Furthermore, there is a correlation between $B/T$ and $r_e$ for both classical and pseudo-bulges,
although they seem to follow offset relations. Ellipticals tend to be more extended than bulges and have
lower surface brightness than classical bulges. Note that no difference is seen in Fig. \ref{fig:nbtmur}
when comparing galaxies with different axial ratios. We have also found that systems with larger $n$ and $B/T$
tend to be in more massive galaxies, although the scatter is large.

In Fig. \ref{fig:hregalmagm} we show that, as expected, galaxies with more extended bulges or discs tend to
be more luminous and massive. This is also true for systems with lower axial ratios.
One might also ask if a similar relation is found if one considers the
component mass, rather than the total mass in the galaxy. Although several stellar systems seem to follow
such a relation, there might be exceptions, such as globular clusters
\citep[see][]{BurBenFab97,KisJorBas06,BarMcLHar07}. This is answered in Fig. \ref{fig:scale}. One sees
that such a correlation indeed exits for pseudo-bulges, classical bulges, bars and discs, separately,
regardless of galaxy axial ratio.
This figure also shows fits to the data, where scale-lengths go $\propto M^\alpha$, where
$\alpha$ is a constant that defines the slope of the relation and $M$ is the component stellar mass. The values
we have found for $\alpha$ for bars, discs, pseudo-bulges, classical bulges and ellipticals are,
respectively, 0.21, 0.33, 0.20, 0.30 and 0.38, with a mean $1\sigma$ uncertainty of 0.02; thus
$\alpha$ seems to increase for more massive components.
It is interesting to note that bars and pseudo-bulges share a very similar value of $\alpha$
(and these are the only components that do so), {\em which is
different from that of classical bulges at a $5\sigma$ level}. This not only suggests a different formation
mechanism for classical and pseudo-bulges, but also reveal a close connection between pseudo-bulges and
bars. Such a connection is expected if pseudo-bulges indeed form through disc instabilities.
Nonetheless, despite such difference in slope, the locus occupied by classical bulges in such
diagram continues rather smoothly from the locus of pseudo-bulges. This suggests that although there
might be two different {\em dominant} formation processes for classical and pseudo-bulges there can be
a significant number of cases where the different processes are at play concomitantly. Thus, the sequence
drawn by pseudo-bulges and classical bulges could represent a sequence where the dominant process
changes rather smoothly from one end to the other.

Another striking feature in Fig. \ref{fig:scale} concerns ellipticals. They are more offset from classical
bulges than classical bulges are offset from pseudo-bulges, as 2D Kolmogorov-Smirnov tests indicate. In fact,
one sees that most classical bulges can not be considered as ellipticals that happen to be surrounded by
a disc, at least not ellipticals more massive than $10^{10}~{\rm M}_\odot$. How robust this result is
against systematic effects from the decompositions? We explore now two possibilities. The first is that
some of the ellipticals could actually be lenticulars whose discs were not identified in the preparation
of the fits. As mentioned in Sect. \ref{sec:method}, we produced test fits in which a disc component was
forced in 20 ellipticals, and found that in only one case the inclusion of a disc could be justified. However,
such a disc contributes with only a few per cent of the total galaxy luminosity, and the resulting shift
in $r_e$, towards lower values, is only of the order of a few hundred parsecs. The second possibility is
that the input values given to the code to fit $r_e$ are systematically larger for ellipticals. We have thus
performed yet another test, again producing test fits for 20 ellipticals but this time with the
input values for $r_e$ cut by half. In most cases, the result was similar to the original fit
(which incidently attests the robustness of the code). In a few cases, the shift in the final value for $r_e$
is again only a few hundred parsecs towards lower values, and in all such cases the resulting $\chi^2$ is
substantially worse than that of the original fit. We thus conclude that such systematic effects
could in principle bring some of the low-mass ellipticals closer to classical bulges in Fig. \ref{fig:scale},
but the result that at least a significant fraction of classical bulges can not be considered as (massive)
ellipticals surrounded by discs seems to be robust. Note that \citet{FisDro08} argue that classical bulges
resemble {\em low-luminosity} ellipticals.

\begin{figure}
   \centering
   \includegraphics[keepaspectratio=true,width=8cm,clip=true]{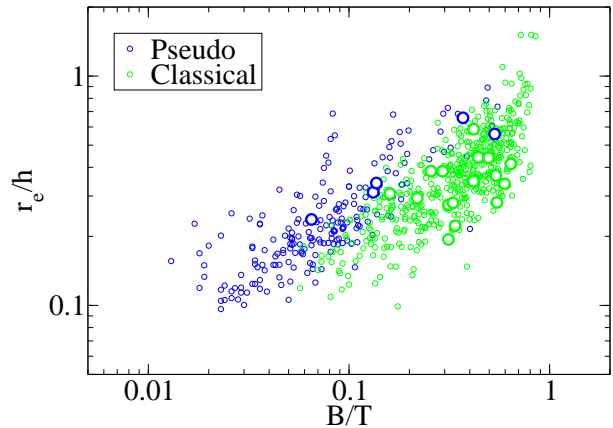}
   \caption{Ratio between the scale-lengths of bulges and discs
$r_e/h$ plotted against $B/T$ for pseudo-bulges and classical bulges. Although
there is some scatter, it is evident that both quantities are correlated, which rules out
the concept of a scale-free Hubble sequence. Interestingly, pseudo-bulges and classical bulges
seem to follow offset relations. For the same $B/T$, pseudo-bulges have on average higher values of
$r_e/h$. Bigger, white-filled circles with similar colour coding are pseudo-bulges and classical bulges
in galaxies with $0.5<b/a<0.7$, and portray similar patterns.}
   \label{fig:rehbt}
\end{figure}

With these data, we can test previous claims that the Hubble sequence is scale-free, i.e. that the
relation between the scale-lengths of bulges and discs does not vary with Hubble type \citep{CoudeJBro96}.
In this context, \citet{GraPri99} showed that fixing the bulge S\'ersic
index, when fitting galaxy surface brightness profiles, can lead to an overestimate or an
underestimate of $r_e$, depending on the true value of $n$. Since $n$ is a free parameter in our
decompositions, our results are not subject to such systematic biases. In Fig. \ref{fig:rehbt},
we plot $r_e/h$ against $B/T$ for pseudo-bulges and classical bulges. This figure shows clearly
that $r_e/h$ and $B/T$ are correlated, and thus the Hubble sequence is not scale-free,
if it is a $B/T$ sequence.
\citet{BalGraPel07} and \citet{LauSalBut07} reached similar results, though with much smaller
samples. Figure \ref{fig:rehbt} also indicates that the relations for pseudo-bulges and classical bulges
are offset: for a constant $B/T$, pseudo-bulges have on average higher values of $r_e/h$.

\begin{figure}
   \centering
   \includegraphics[keepaspectratio=true,width=8cm,clip=true]{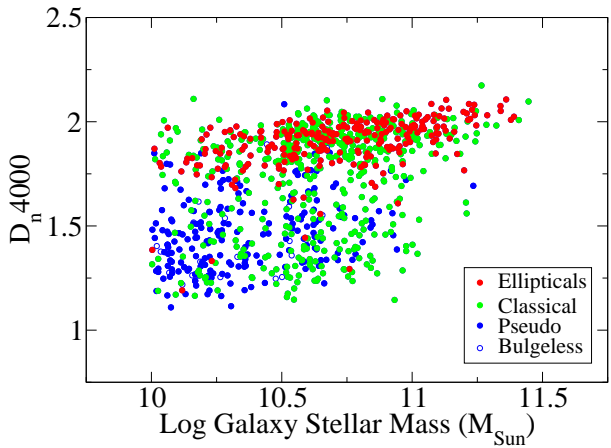}
   \caption{The 4000\AA\, break index plotted against galaxy dust-corrected stellar mass. Virtually
all elliptical galaxies are found in the red sequence, while practically all galaxies with pseudo-bulges,
and bulgeless galaxies, are found in the blue cloud. Galaxies with classical bulges can be found in both
the red sequence and the blue cloud, although the majority of them occupy the red sequence.}
   \label{fig:d4mas}
\end{figure}

It is well known that galaxies can be separated depending on the locus they occupy in a
colour-magnitude diagram. Some galaxies follow a tight colour-magnitude relation, in the sense
that more luminous galaxies are redder, known as the red sequence, whereas others spread beside
the red sequence, originating a swarm of points known as the blue cloud
\citep[see e.g.][and references therein]{GlaLopYee98,BelZhePap07}. Such a diagram is a powerful
tool, able to provide clues on the formation and evolutionary processes of galaxies
\citep[see e.g.][]{DeLPogAra07}. \citet{DroFis07}, based on a careful analysis of a sample of
39 galaxies, suggested that galaxies with pseudo-bulges populate the blue cloud,
while galaxies with classical bulges populate
the red sequence. But colour can be subject to dust attenuation effects, and galaxy magnitude
is actually used as a proxy to galaxy stellar mass. Since we have measurements of D$_n$(4000),
an efficient, dust-insensitive index to identify stellar populations with different mean ages, as well as
carefully modelled stellar masses, we can go a step further and plot D$_n$(4000) against
galaxy stellar mass, as is done in Fig. \ref{fig:d4mas}. It should be noted that often one uses
the {\em total} galaxy colour in the colour-magnitude diagram, whereas our D$_n$(4000) values correspond
to the {\em central} galaxy colour. That is, however, an asset, since, as argued in Sect.
\ref{sec:pseudo}, the light that goes through the SDSS fibre, from which D$_n$(4000) is measured, is
in our sample {\em mostly} emitted by stars in the bulge, meaning that our Fig. \ref{fig:d4mas} is mainly
a comparison between ellipticals, classical bulges and pseudo-bulges, avoiding in most cases effects resulting
from disc light. Nonetheless, we have verified that similar results emerge if we plot the colour-magnitude
diagram in its usual form, using dereddened $g-i$ Petrosian colours from the SDSS database and absolute
magnitudes from \citet{KauHecWhi03}. In such case, however, the blue cloud is not as detached from
the red sequence as in Fig. \ref{fig:d4mas}, most likely due to dust attenuation.
Our results show clearly that virtually
all elliptical galaxies are found in the red sequence, and practically all galaxies with
pseudo-bulges, and bulgeless galaxies, are found in the blue cloud. However, although
most of the galaxies with classical bulges do indeed populate the red sequence, a significant
fraction of them are found in the blue cloud (see also Fig. \ref{fig:d4col}).

\subsection{The fundamental plane}
\label{sec:FP}

Elliptical galaxies follow well-defined relations between their effective radius $r_e$, mean effective
surface brightness $\left<\mu_e\right>$
and central velocity dispersion $\sigma_0$, known as the fundamental plane (FP). Furthermore,
it is possible to show that such relations arise in virialised systems
\citep{DjoDav87,DreLynBur87,BenBurFab92}. Thus, the FP provides important clues
on the formation processes of stellar systems. \citet{BenBurFab92} proposed a formulation
of the FP, defining a space, whose three axes, $\kappa_1$, $\kappa_2$ and $\kappa_3$, are
defined in terms of $r_e$, $\left<\mu_e\right>$ and $\sigma_0$, and are proportional to important
physical parameters: $\kappa_1$ is proportional to the logarithm of the mass,
$\kappa_2$ is proportional mainly to the logarithm of the surface brightness, and $\kappa_3$ is
proportional to the logarithm of the mass-to-light ratio. Here, mass means {\em dynamical}
mass, i.e. stellar mass plus the dark matter mass content.

What is the locus occupied by pseudo-bulges and classical bulges in the FP? We can answer this question
using $r_e$ and $\left<\mu_e\right>$ from our decompositions, and velocity dispersion measurements
from the SDSS database. We have chosen to use the estimates provided in the Data Release 6 (DR6), which
are based on results from a direct fit to the spectra. In earlier releases, the estimates are an
average of results from direct-fitting and Fourier-fitting, but the latter systematically overestimates the velocity
dispersion for low-mass galaxies \citep[their Fig. 17]{BerSheAnn03a}. In fact, we have verified that the velocity
dispersion estimates from DR2 are systematically larger (by $\sim10-20$ per cent)
than those from DR6, for the galaxies in our sample, when the velocity dispersion is below $\approx150$ km/s.
Due to the limited spectral resolution of the SDSS spectra, it is recommended to use only spectra with
signal-to-noise ratio above 10, and without warning flags. We have verified that only one galaxy in our sample
does not comply with the former criterion, and only 8 do not comply with the latter.
Before we can use these measurements we have to apply aperture corrections
due to the fact that these velocity dispersions are obtained through a fixed fibre aperture of 3 arcsec,
which is physically different from galaxy to galaxy, depending on their distances. We thus used
the prescription provided by \citet{JorFraKja95} and converted the SDSS velocity dispersion measurements
to the velocity dispersion at 1/8 of $r_e$, $\sigma_8$. This empirical prescription is based on
measurements for elliptical and lenticular galaxies, and it is unclear if it is also valid for disc-dominated
galaxies. However, the corrections applied are small, and do not affect our results significantly.

\begin{figure*}
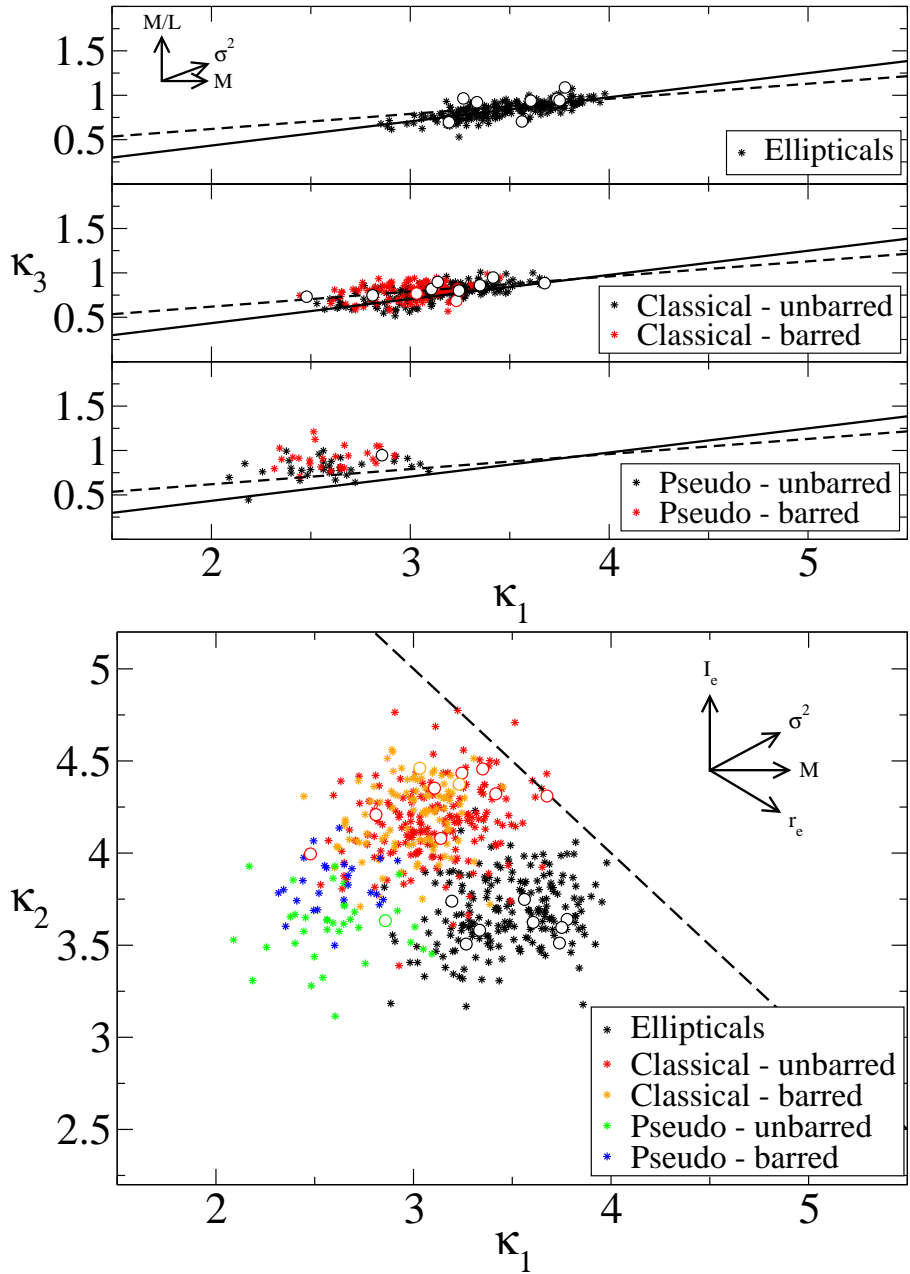

   \centering
   \includegraphics[keepaspectratio=true,width=12cm,clip=true]{kappa1_dr6_new.eps}
   \includegraphics[keepaspectratio=true,width=12cm,clip=true]{kappa2_dr6_new.eps}
   \caption{Elliptical galaxies, classical bulges and pseudo-bulges in the $\kappa$-space
formulation of the fundamental plane (FP). The top three panels show the edge-on view
of the FP, $\kappa_3$ plotted against $\kappa_1$, while the bottom panel shows its nearly face-on
view, $\kappa_2$ plotted against $\kappa_1$. Barred and unbarred galaxies are also indicated.
In the top panels, the solid line is a fit to our ellipticals, while the dashed line is the fit
obtained by \citet{BerSheAnn03c} for nearly 9000 early-type galaxies. The dashed line in the bottom
panel shows the limit of the zone of avoidance. Arrows indicate how some important physical parameters
vary accross the FP. Bigger, white-filled circles represent systems with $0.5<b/a<0.7$, with similar
colour coding.}
   \label{fig:kappa}
\end{figure*}

We thus calculated $\kappa_1$, $\kappa_2$
and $\kappa_3$ for elliptical galaxies, classical bulges and pseudo-bulges,
using the definitions in \citet{BenBurFab92}, and applying an offset in surface
brightness to facilitate comparison with \citet{BenBurFab92} and \citet{BurBenFab97}, who used
measurements in the $B$-band (the same procedure was done also by \citealt{BerSheAnn03c}).
The top panels of Fig. \ref{fig:kappa} show that the relation we have found for
$\kappa_3$ against $\kappa_1$, considering only elliptical galaxies, is similar to that found
by \citet{BerSheAnn03c}, which used SDSS $i$-band data for a sample of nearly 9000 early-type galaxies.
\citet{BerSheAnn03c} used estimates of the velocity dispersion that include Fourier-fitting results,
which, as mentioned above, overestimate the velocity dispersion for low-mass galaxies. Accordingly,
their relation is somewhat less steep than ours, even though they have excluded estimates below 100 km/s.
We find that

\begin{equation}
\kappa_3=0.27\kappa_1-0.11.
\end{equation}

\noindent In addition, classical bulges also seem to follow the same relation, although somewhat offset
to higher $\kappa_3$ values, regardless of the galaxy being
barred or unbarred \citep[see also][]{FalPelBal02}. Classical bulges follow more closely the relation
found by \citet{BerSheAnn03c}. It should be noted that their sample also includes
early-type {\em disc} galaxies. To select their sample of early-type galaxies they have
included only galaxies with concentration index $R90/R50$ larger than 2.5.
This does not mean, however,
that their sample contains only elliptical galaxies. In fact, we have checked that 99 per cent of our ellipticals
have $R90/R50>2.5$, but this is also the case for 76 per cent of our galaxies with classical bulges, and 8
per cent of our galaxies with pseudo-bulges. \citet{BerSheAnn03a} also used other criteria to avoid including
late-type galaxies in their sample, but it is unlikely that these criteria excluded most of the galaxies with
classical bulges.
This can also explain why our relation for ellipticals is slightly different from theirs.

\begin{figure*}
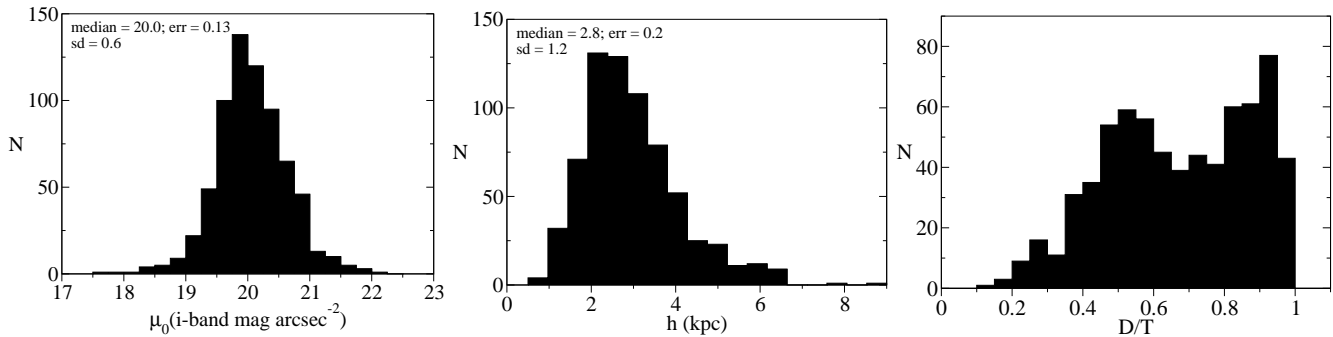

   \centering
   \includegraphics[width=5.8cm,clip=true,keepaspectratio=true]{mu_0.eps}
   \includegraphics[width=5.8cm,clip=true,keepaspectratio=true]{h.eps}
   \includegraphics[keepaspectratio=true,width=5.8cm,clip=true]{dt.eps}
   \caption{Distributions of $\mu_0$, $h$ and $D/T$ for all disc galaxies in the sample. The first two panels
also show the values of the median and standard deviation of each distribution, as well as the
mean 1$\sigma$ uncertainty in individual measurements. Bin sizes are $\approx2\sigma$.}
   \label{fig:dispardis}
\end{figure*}

Pseudo-bulges in both barred and unbarred galaxies lie significantly and
systematically above these relations. This offset can not be attributed to effects due to
dust attenuation alone.
Even considering the upper limit for dust attenuation discussed in Sect. \ref{sec:pseudo},
such a correction will bring the points down vertically by 0.19 in $\kappa_3$, while one sees that
the offset is $\approx0.3$, on average. In this projection, 2D Kolmogorov-Smirnov tests indicate that
pseudo-bulges differ more from classical bulges than ellipticals differ from classical bulges or bulges
altogether. Pseudo-bulges also lie in a locus in the $\kappa_2\times\kappa_1$
projection of the FP which is different from those of both ellipticals and classical bulges
(bottom panel of Fig. \ref{fig:kappa}). Again, correcting for upper limit
dust effects would put pseudo-bulges
up along the $\kappa_2$ axis only by 0.27. In this projection, the statistical tests indicate that
pseudo-bulges differ from classical bulges as much as ellipticals differ from classical bulges or bulges
altogether. In both projections, pseudo-bulges lie in regions close
to those occupied by disc-dominated galaxies \citep[see][]{PieGavFra02}.
We note that barred galaxies tend to be more separated from their unbarred counterparts in the case
of galaxies with pseudo-bulges, as compared to galaxies with classical bulges.
Barred galaxies seem to have slightly higher values of both $\kappa_2$ and $\kappa_3$, on average.
This is also confirmed with the statistical tests.
Finally, it is also worth noting that, as expected from the results in Sect. \ref{sec:scale},
systems with lower axial ratios seem to generally share the same properties as those with
$b/a\geq0.9$.

\subsection{Discs}
\label{sec:discs}

\begin{figure}
   \centering
   \includegraphics[keepaspectratio=true,width=8cm,clip=true]{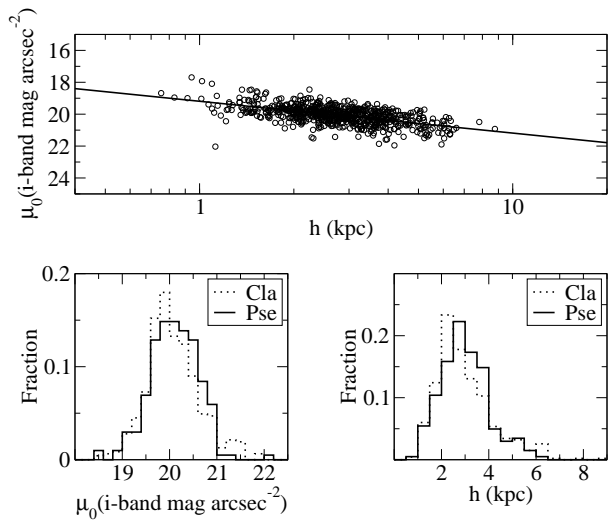}
   \caption{Scaling relation between disc scale-length and central surface brightness. The histograms
at the bottom show that galaxies with pseudo-bulges have more extended discs, with fainter
central surface brightness, on average, as compared with galaxies hosting classical bulges, albeit with significant
overlap. A Kolmogorov-Smirnov test rejects the null hypothesis of no difference between such distributions
at a 92 per cent confidence level.}
   \label{fig:muh}
\end{figure}

Figure \ref{fig:dispardis} shows the distributions of the disc parameters ($\mu_0$, $h$ and $D/T$).
It is worth noticing that the distribution of disc-to-total ratio shows two peaks which match
the peaks in the bulge-to-total ratio distributions for classical bulges and pseudo-bulges
(shown in Fig. \ref{fig:car99}).
The two peaks in the distribution of $D/T$ are at about 0.5 and 0.9.
Bars account typically for 10 per cent of the galaxy luminosity.
In Fig. \ref{fig:muh} we show the scaling relation between disc scale-length and central surface
brightness. When one separates galaxies by their bulge type (classical vs. pseudo), one sees that,
on average, galaxies with pseudo-bulges have more extended discs, with fainter
central surface brightness, as compared with galaxies hosting classical bulges, albeit with significant
overlap. A Kolmogorov-Smirnov test rejects the null hypothesis of no difference between such distributions
at a 92 per cent confidence level. It is also
interesting to ask how the stellar mass in discs varies between these two
galaxy classes. The stellar mass in discs of galaxies with pseudo-bulges is, on average,
$1.15\times10^{10}~{\rm M}_\odot$, whereas that in discs of galaxies with classical bulges
is $1.93\times10^{10}~{\rm M}_\odot$, thus a factor of 1.7 larger. A similar factor (1.8)
corresponds to bars, when present. We can also ask how do their
bulge masses compare. The stellar mass within pseudo-bulges is typically
$2.2\times10^{9}~{\rm M}_\odot$, whereas in classical bulges this is $1.41\times10^{10}~{\rm M}_\odot$
(see Table \ref{tab:masdis}), thus a factor of 6.4 larger. Hence, in terms of stellar mass,
there is more similarity between discs (and bars) in
galaxies with pseudo-bulges and those in galaxies with classical bulges, than between classical and
pseudo-bulges themselves. Furthermore, the mean difference between the total galaxy mass of
galaxies hosting pseudo-bulges and of those with classical bulges is a factor of 2. Therefore, if one scales
down a typical galaxy with a classical bulge, dividing the mass of each of its components by 2,
one ends up with a galaxy that has a disc (and a bar) with mass similar to that of the corresponding component
in galaxies with pseudo-bulges. However, the bulge of such scaled-down galaxy is around 3 times more massive
than the typical pseudo-bulge. If we assume that the physical processes that lead to the formation of discs
and bars are the same regardless of the resulting bulge, then this difference suggests that the formation
processes of classical bulges and pseudo-bulges are typically distinct.

\subsection{A compendium of galaxy structural parameters}
\label{sec:comp}

\begin{table*}
\centering
\begin{minipage}{171mm}
\caption{Structural parameters of bulges, discs and bars in our sample from $i$-band image fits. Galaxy
identifications, uncertainties, and similar tables in the $g$ and $r$ bands
are also available. The printed version of the paper contains only a sample. The full table is available
in the electronic issue of the journal.}
\scriptsize
\label{tab:res}
\begin{tabular}{@{}ccccccccccccccccc@{}}
\hline \hline
$\mu_0$ & $h$ & $\mu_e$ & $r_e$ & $n$ & $\mu_{e,{\rm bar}}$ & $r_{e,{\rm bar}}$ &
$\epsilon_{\rm bar}$ & $n_{\rm bar}$ & $L_{\rm bar}$ & $c$ & $B/T$ & $D/T$ & $Bar/T$ & $\chi^2$ &
seeing$_{\rm HWHM}$ & $z$ \\
(1) & (2) & (3) & (4) & (5) & (6) & (7) & (8) & (9) & (10) & (11) & (12) & (13) & (14) & (15) &
(16) & (17) \\
\hline
19.41 &  2.10 &  19.72  & 1.24 &  1.29 &  0.0    &   0.0   &   0.0   &    0.0  &  0.0   &  0.0    &     0.35 &  0.64  & 0.0  &  1.14 &  1.32   &    0.067  \\
19.16 &  2.40 &  19.89  & 1.82 &  3.54 &  0.0    &   0.0   &   0.0   &    0.0  &  0.0   &  0.0    &     0.48 &  0.51  & 0.0  &  1.42 &  0.68   &    0.047  \\
19.25 &  2.18 &  20.54  & 0.52 &  0.94 &  0.0    &   0.0   &   0.0   &    0.0  &  0.0   &  0.0    &     0.03 &  0.96  & 0.0  &  1.65 &  0.45   &    0.043  \\
0.0   &  0.0  &  20.19  & 2.73 &  2.57 &  0.0    &   0.0   &   0.0   &    0.0  &  0.0   &  0.0    &     1.0  &  0.0   & 0.0  &  1.03 &  0.91   &    0.063  \\
0.0   &  0.0  &  20.46  & 3.12 &  4.32 &  0.0    &   0.0   &   0.0   &    0.0  &  0.0   &  0.0    &     1.0  &  0.0   & 0.0  &  1.02 &  0.82   &    0.056  \\
18.95 &  2.42 &  20.89  & 1.33 &  0.82 &  0.0    &   0.0   &   0.0   &    0.0  &  0.0   &  0.0    &     0.08 &  0.91  & 0.0  &  1.66 &  0.94   &    0.065  \\
0.0   &  0.0  &  21.23  & 2.32 &  4.53 &  0.0    &   0.0   &   0.0   &    0.0  &  0.0   &  0.0    &     1.0  &  0.0   & 0.0  &  1.18 &  0.67   &    0.046  \\
0.0   &  0.0  &  19.69  & 1.63 &  5.15 &  0.0    &   0.0   &   0.0   &    0.0  &  0.0   &  0.0    &     1.0  &  0.0   & 0.0  &  1.56 &  0.94   &    0.044  \\
0.0   &  0.0  &  19.97  & 2.08 &  3.78 &  0.0    &   0.0   &   0.0   &    0.0  &  0.0   &  0.0    &     1.0  &  0.0   & 0.0  &  1.61 &  0.65   &    0.045  \\
0.0   &  0.0  &  19.55  & 1.36 &  4.50 &  0.0    &   0.0   &   0.0   &    0.0  &  0.0   &  0.0    &     1.0  &  0.0   & 0.0  &  1.29 &  0.68   &    0.047  \\
20.03 &  2.59 &  21.52  & 0.65 &  1.72 &  20.92  &   1.39  &   0.69  &    0.51 &  2.11  &  3.0    &     0.02 &  0.92  & 0.04 &  1.67 &  0.55   &    0.047  \\
0.0   &  0.0  &  20.64  & 2.03 &  3.12 &  0.0    &   0.0   &   0.0   &    0.0  &  0.0   &  0.0    &     1.0  &  0.0   & 0.0  &  1.83 &  1.00   &    0.047  \\
21.41 &  3.21 &  20.22  & 1.45 &  6.48 &  0.0    &   0.0   &   0.0   &    0.0  &  0.0   &  0.0    &     0.69 &  0.31  & 0.0  &  1.30 &  0.56   &    0.047  \\
0.0   &  0.0  &  21.05  & 1.98 &  2.47 &  0.0    &   0.0   &   0.0   &    0.0  &  0.0   &  0.0    &     1.0  &  0.0   & 0.0  &  1.24 &  0.51   &    0.042  \\
0.0   &  0.0  &  20.86  & 2.38 &  4.01 &  0.0    &   0.0   &   0.0   &    0.0  &  0.0   &  0.0    &     1.0  &  0.0   & 0.0  &  1.28 &  0.66   &    0.046  \\
20.17 &  1.84 &  20.94  & 2.02 &  4.18 &  20.56  &   1.73  &   0.63  &    0.74 &  5.08  &  2.4    &     0.58 &  0.29  & 0.12 &  1.54 &  0.61   &    0.042  \\
20.78 &  3.25 &  21.36  & 0.58 &  1.17 &  0.0    &   0.0   &   0.0   &    0.0  &  0.0   &  0.0    &     0.03 &  0.96  & 0.0  &  0.91 &  0.71   &    0.049  \\
0.0   &  0.0  &  20.16  & 1.48 &  3.17 &  0.0    &   0.0   &   0.0   &    0.0  &  0.0   &  0.0    &     1.0  &  0.0   & 0.0  &  1.53 &  0.55   &    0.047  \\
20.80 &  3.91 &  21.07  & 2.11 &  3.01 &  20.01  &   1.52  &   0.62  &    0.86 &  4.15  &  2.36   &     0.0  &  0.0   & 0.0  &  1.76 &  0.95   &    0.046  \\
20.33 &  2.6  &  21.66  & 1.77 &  3.88 &  0.0    &   0.0   &   0.0   &    0.0  &  0.0   &  0.0    &     0.34 &  0.65  & 0.0  &  2.02 &  0.94   &    0.045  \\
\hline
\end{tabular}
Columns (1) and (2) show the disc central surface brightness and scale-length, respectively. Columns (3), (4) and
(5) give, respectively, bulge effective surface brightness, effective radius and S\'ersic index. Bar parameters
are given in columns (6) to (11), respectively: effective surface brightness, effective radius,
ellipticity, S\'ersic index, semi-major axis and boxyness. Bulge-to-total, disc-to-total and bar-to-total
luminosity ratios are given in columns (12), (13) and (14), respectively. The $\chi^2$ value of the fit
is shown in column (15). Finally, columns (16) and (17), show, respectively, the seeing HWHM of the
observation and the galaxy redshift. Intensities are in units of mag arcsec$^{-2}$ and lengths in kpc.
Seeing is in arcsec.
\end{minipage}
\normalsize
\end{table*}

\begin{table*}
\centering
\begin{minipage}{170mm}
\caption{Absolute magnitudes and integrated colours of bulges, discs and bars in our sample from the {\sc budda}
models. The printed version of the paper contains only a sample. The full table is available in the electronic
issue of the journal.}
\scriptsize
\label{tab:totmagcol}
\begin{tabular}{@{}ccccccc@{\hspace{2mm}}c@{\hspace{2mm}}c@{\hspace{2mm}}c@{\hspace{2mm}}c@{\hspace{2mm}}c@{\hspace{2mm}}c@{\hspace{2mm}}c@{\hspace{2mm}}c@{}}
\hline \hline
M$_{\rm d}$($g$) & M$_{\rm d}$($r$) & M$_{\rm d}$($i$) & M$_{\rm b}$($g$) &
M$_{\rm b}$($r$) & M$_{\rm b}$($i$) & M$_{\rm bar}$($g$) & M$_{\rm bar}$($r$) & M$_{\rm bar}$($i$) &
$(g-i)_{\rm d}$ & $(r-i)_{\rm d}$ & $(g-i)_{\rm b}$ & $(r-i)_{\rm b}$ & $(g-i)_{\rm bar}$ & $(r-i)_{\rm bar}$ \\
(1) & (2) & (3) & (4) & (5) & (6) & (7) & (8) & (9) & (10) & (11) & (12) & (13) & (14) & (15) \\
\hline
-19.74 &  -20.57 &  -20.96 &  -19.33 &  -19.93 &  -20.33 &  0.0     & 0.0     & 0.0     & 1.22  &  0.38  &  1.00  &  0.39  &  0.0    &   0.0  \\
-20.17 &  -20.87 &  -21.34 &  -19.40 &  -20.68 &  -21.26 &  0.0     & 0.0     & 0.0     & 1.17  &  0.46  &  1.86  &  0.58  &  0.0    &   0.0  \\
-20.17 &  -20.64 &  -20.93 &  -15.81 &  -16.86 &  -17.23 &  0.0     & 0.0     & 0.0     & 0.76  &  0.29  &  1.42  &  0.37  &  0.0    &   0.0  \\
0.0    &  0.0    &  0.0    &  -20.67 &  -21.39 &  -21.73 &  0.0     & 0.0     & 0.0     & 0.0   &  0.0   &  1.06  &  0.34  &  0.0    &   0.0  \\
0.0    &  0.0    &  0.0    &  -20.61 &  -21.37 &  -21.82 &  0.0     & 0.0     & 0.0     & 0.0   &  0.0   &  1.21  &  0.45  &  0.0    &   0.0  \\
-20.69 &  -21.25 &  -21.54 &  -15.50 &  -18.43 &  -18.97 &  0.0     & 0.0     & 0.0     & 0.84  &  0.29  &  3.47  &  0.53  &  0.0    &   0.0  \\
0.0    &  0.0    &  0.0    &  -19.17 &  -20.13 &  -20.53 &  0.0     & 0.0     & 0.0     & 0.0   &  0.0   &  1.35  &  0.4   &  0.0    &   0.0  \\
0.0    &  0.0    &  0.0    &  -19.86 &  -20.79 &  -21.33 &  0.0     & 0.0     & 0.0     & 0.0   &  0.0   &  1.46  &  0.53  &  0.0    &   0.0  \\
0.0    &  0.0    &  0.0    &  -20.15 &  -20.99 &  -21.42 &  0.0     & 0.0     & 0.0     & 0.0   &  0.0   &  1.27  &  0.43  &  0.0    &   0.0  \\
0.0    &  0.0    &  0.0    &  -19.50 &  -20.51 &  -20.95 &  0.0     & 0.0     & 0.0     & 0.0   &  0.0   &  1.44  &  0.43  &  0.0    &   0.0  \\
-19.89 &  -20.34 &  -20.61 &  -14.60 &  -16.17 &  -16.75 &  -16.76  & -17.07  & -17.42  & 0.71  &  0.26  &  2.15  &  0.57  &  0.66   &   0.34 \\
0.0    &  0.0    &  0.0    &  -19.25 &  -20.10 &  -20.67 &  0.0     & 0.0     & 0.0     & 0.0   &  0.0   &  1.42  &  0.56  &  0.0    &   0.0  \\
-18.31 &  -19.23 &  -19.64 &  -19.26 &  -20.12 &  -20.51 &  0.0     & 0.0     & 0.0     & 1.32  &  0.41  &  1.24  &  0.39  &  0.0    &   0.0  \\
0.0    &  0.0    &  0.0    &  -18.99 &  -19.80 &  -20.15 &  0.0     & 0.0     & 0.0     & 0.0   &  0.0   &  1.15  &  0.35  &  0.0    &   0.0  \\
0.0    &  0.0    &  0.0    &  -19.66 &  -20.31 &  -20.70 &  0.0     & 0.0     & 0.0     & 0.0   &  0.0   &  1.03  &  0.38  &  0.0    &   0.0  \\
-18.84 &  -19.48 &  -19.80 &  -19.46 &  -20.22 &  -20.53 &  -17.72  & -18.74  & -18.82  & 0.96  &  0.32  &  1.07  &  0.30  &  1.09   &   0.07 \\
-19.67 &  -20.16 &  -20.39 &  -15.23 &  -16.32 &  -16.74 &  0.0     & 0.0     & 0.0     & 0.72  &  0.23  &  1.51  &  0.42  &  0.0    &   0.0  \\
0.0    &  0.0    &  0.0    &  -19.17 &  -20.01 &  -20.40 &  0.0     & 0.0     & 0.0     & 0.0   &  0.0   &  1.23  &  0.38  &  0.0    &   0.0  \\
-19.41 &  -20.19 &  -20.68 &  -19.33 &  -20.35 &  -20.62 &  -17.87  & -18.82  & -19.14  & 1.27  &  0.49  &  1.28  &  0.26  &  1.27   &   0.32 \\
-19.39 &  -20.19 &  -20.34 &  -19.01 &  -17.96 &  -19.66 &  0.0     & 0.0     & 0.0     & 0.94  &  0.14  &  0.65  &  1.70  &  0.0    &   0.0  \\
\hline
\end{tabular}
Columns (1) to (3) show the disc absolute magnitude in the $g$, $r$ and $i$ bands, respectively. Columns (4) to (9)
show the same parameters for bulge and bar, respectively, as indicated. Columns (10) and (11), (12) and (13), and (14)
and (15) give the integrated $(g-i)$ and $(r-i)$ colours of disc, bulge and bar, respectively, as indicated.
\end{minipage}
\normalsize
\end{table*}

The full results from our decompositions are given in Tables \ref{tab:res} and \ref{tab:totmagcol}.
Table \ref{tab:bulpardis} shows the parameters describing the distributions (median and standard
deviation), as well as the mean 1$\sigma$ uncertainty in individual measurements, of $\mu_e$, $r_e$
and $n$ for elliptical galaxies, classical bulges and pseudo-bulges.

\begin{table*}
\centering
\begin{minipage}{132mm}
\caption{Parameters describing the distributions (median and standard deviation), as well as the mean
1$\sigma$ uncertainty in individual measurements, of $\mu_e$, $r_e$ and $n$ for elliptical galaxies,
classical bulges and pseudo-bulges from the $i$-band decompositions.}
\label{tab:bulpardis}
\begin{tabular}{@{}lccccccccc@{}}
\hline \hline
\omit & \omit & $\mu_e$ (mag arcsec$^{-2}$)& \omit & \omit & $r_e$ (kpc) & \omit & \omit & $n$ & \omit \\
\omit & median & sd & 1$\sigma$ & median & sd & 1$\sigma$ & median & sd & 1$\sigma$ \\
\hline
Elliptical & 20.8 & 0.4 & 0.10 & 3.0 & 1.0 & 0.15 & 3.8 & 0.9 & 0.5 \\
Classical  & 19.6 & 0.4 & 0.05 & 1.0 & 0.4 & 0.03 & 3.4 & 1.3 & 0.4 \\
Pseudo     & 20.5 & 0.5 & 0.07 & 0.7 & 0.3 & 0.03 & 1.5 & 0.9 & 0.2 \\
\hline
\end{tabular}
\end{minipage}
\end{table*}

Evidently, theoretical work on galaxy formation and evolution should comply with these results, either
by predicting compatible structural properties, or by using these results as
a starting point to build realistic
models. The same thing can be said about the masses of the different components,
and their mass distributions, which are presented in the next section.
These results are also useful to test automated procedures of bulge/disc/bar 2D decomposition, with
no checking or direct human intervention at the decomposition of every single object. An automated code
able to reproduce these results using the same input data (i.e. the same images) should be deemed successful
and reliable to be used with much larger samples, which are lately becoming commonly available.
The results from our image decompositions are available for download.\footnote{See
http://www.mpa-garching.mpg.de/$\sim$dimitri/buddaonsdss/\linebreak buddaonsdss.html .} These include all
structural parameters of bulges, discs and bars in the $g$, $r$ and $i$ bands, absolute magnitudes
and integrated $g-i$ and $r-i$ colours of each component separately, as well as model images for the whole
galaxy and for each component separately.

\subsection{The stellar mass and luminosity budgets at $z\sim0$}
\label{sec:budget}

\begin{table}
\centering
\begin{minipage}{64mm}
\caption{Parameters describing the distributions (median and standard deviation) of the stellar mass
in elliptical galaxies, classical bulges, pseudo-bulges, discs and bars.}
\label{tab:masdis}
\begin{tabular}{@{}lcc@{}}
\hline \hline
\omit & median ($10^{10}~{\rm M}_\odot$) & sd ($10^{10}~{\rm M}_\odot$) \\
\hline
Ellipticals & 5.15 & 4.37 \\
Classical   & 1.41 & 1.98 \\
Pseudo      & 0.22 & 0.99 \\
Discs       & 1.66 & 1.62 \\
Bars        & 0.40 & 0.61 \\
\hline
\end{tabular}
\end{minipage}
\end{table}

With the $g-i$ integrated colour of each galaxy component, we have obtained the
corresponding mass-to-light ratio in the $i$-band, using the results found by
\citet[see their Fig. 15]{KauHecBud07}. They showed that the mass-to-light ratio in the $i$-band
is well correlated with $g-i$ colour over a range of more than a factor 10 in mass-to-light ratio and
with a scatter of less than 0.1 dex. They found that the $g-i$ colour is the colour that shows the
tightest relation with mass-to-light ratio. Furthermore, \citet{BeldeJ01} showed that the combined
effects of stellar age and dust attenuation are such that the stellar masses obtained here with
these mass-to-light ratios are not significantly affected by dust.

With the mass-to-light ratio and the total luminosity of each component we were thus able to calculate
the stellar mass content in each component separately. Interestingly, we have verified that the
total mass in each {\em galaxy}, calculated by adding the masses of the individual galaxy
components, agrees very well with the galaxy stellar mass estimated in \citet{KauHecWhi03}, with
a scatter of less than 0.02 dex. The parameters that characterise the stellar mass distributions of
each galaxy component (median and standard deviation) are shown in Table \ref{tab:masdis}.
All such distributions are skewed towards lower mass values.

We were also able to compute the fraction of the total stellar mass content in our galaxies contained
within each component separately, thus obtaining the stellar mass budget in the local universe
for massive galaxies (more massive than $10^{10}~{\rm M}_\odot$). The following figures take into account
the selection effect due to our axial ratio cut
discussed in Sect. \ref{sec:sample}. Elliptical galaxies contain
32 per cent of the total stellar mass, and the corresponding values for discs, bulges and bars are,
respectively, 36, 28 and 4 per cent. Classical bulges contain 25 per cent of the total stellar mass,
while pseudo-bulges contain 3 per cent. Hence, 11 per cent of the stellar mass in bulges are in
pseudo-bulges, the remaining being in classical bulges. The uncertainty in these fractions from
Poisson statistics only is between 1 and 2 percentage points.

The corresponding luminosity fractions are as follows: 27 per cent of the total luminosity from
the galaxies in our sample (in the $i$-band) come from elliptical galaxies, 46 per cent from
discs, 24 per cent from bulges, and 3 per cent from bars. Classical bulges emit 20 per cent of the
total luminosity, while pseudo-bulges account for 4 per cent. Hence, 16 per cent of the luminosity from
bulges come from pseudo-bulges, the remaining coming from classical bulges.

As expected, luminosity ratios, such as the bulge-to-total ratio $B/T$, change in different bands.
For pseudo-bulges, the median value of $B/T$ in the $g$-band is 0.053, whereas in the $i$-band
it is 0.075. For classical bulges, the median value of $B/T$ in the $g$-band is 0.349, and in the $i$-band
0.377. The median value of $Bar/T$ also increases from 0.087 in the $g$-band to 0.095 in the
$i$-band. The median value of $D/T$ {\em decreases} from 0.702 in the $g$-band to 0.684 in the
$i$-band.

It is also worth noting the fraction in number corresponding to the different galaxy categories. From the
total number of galaxies in our sample, 22 per cent are ellipticals, 76 per cent are disc
galaxies with bulges, and 2 per cent are bulgeless galaxies (i.e. pure discs).
Within disc galaxies with bulges, 68 per cent host classical bulges, while 32 per cent have
pseudo-bulges. Bars are found in 42 per cent of all disc galaxies.

\section{Discussion}
\label{sec:dis}

\subsection{Comparison with previous work}

\citet{AllDriGra06} also performed parametric decomposition of a large sample
of galaxy images. As mentioned in the Introduction, their sample is significantly larger than ours,
but our methodology aims at a more careful analysis. They found that the stellar mass content in the local
universe is distributed as 13 per cent in elliptical galaxies, 58 per cent in discs, 26 per cent in classical
bulges, 1.5 per cent in pseudo-bulges and the remainder in low-luminosity blue spheroids
\citep[][see also \citealt{BenDzaFre07} who found that $35-51$ per cent of the stellar mass is
in discs]{DriAllLis07}. The figures do not change substantially when dust is taken into account \citep{DriPopTuf07}.
A direct comparison with our results is difficult,
since their sample was not selected with a cut in stellar mass. As we avoided low-mass galaxies, it is natural
to expect that we find a larger fraction of the local stellar mass in ellipticals, and a corresponding
lower fraction in discs. Interestingly, however, their fraction
of stellar mass in classical bulges is similar to what we find.
This is somewhat unexpected, considering the different sample selection criteria,
since their sample should contain a larger
fraction of disc-dominated galaxies than our sample. This might be explained by the absence of a bar component in
their models. As discussed in \citet{Gad08b}, it is possible to obtain good (but wrong) fits for barred galaxies,
with no bar component in the model, because the bulge model absorbs the bar.
Thus, their bulge-to-total luminosity ratios
could be overestimated. They have considered pseudo-bulges as those with $B/T<0.5$ and $r_e/h<0.5$, and found them
to be generally bluer than the remaining bulges.
From our Fig. \ref{fig:rehbt}, one sees that such criteria still include many classical bulges.
However, they found a stellar content in pseudo-bulges smaller than ours, albeit within the
uncertainties. A possible explanation for that could be that a number of their pseudo-bulges were
smoothed out by the PSF. Although they have secured that the galaxies in their sample have effective
radii larger than the PSF HWHM, it was shown in \citet{Gad08b} that it is the effective radius of the {\em bulge}
that has to be considered, and not that of the whole galaxy. This issue is aggravated
in \citet{BenDzaFre07}, who used SDSS images up to a redshift of 0.3.

In order to try a more meaningful comparison with the results in \citet{DriAllLis07} we can assume that
all galaxies with stellar masses below $10^{10}~{\rm M}_\odot$ are pure discs. This certainly underestimates
the stellar mass content in bars and pseudo-bulges, but can give us a reasonable assessment of the distribution
of stellar mass in discs, ellipticals and classical bulges, as if we have made no mass cut.
Using the stellar mass function derived by
\citet{BelMcIKat03}, using $g$-band data for a local sample of SDSS galaxies, we find that 22 per cent
of the stellar mass in the local universe is in galaxies with stellar masses below $10^{10}~{\rm M}_\odot$.
If all this mass is in discs, then our local stellar mass budget becomes as follows. Discs contain 50 per cent
of the stellar mass, ellipticals contain 25 per cent of stellar mass, classical bulges and pseudo-bulges contribute
with, respectively, 20 per cent and 2 per cent, and finally bars contain 3 per cent of the stellar
mass. One thus sees a more reasonable agreement, although we find a somewhat larger mass fraction in ellipticals.

In \citet{Gad08b} it was argued, with indirect means, that the stellar mass content in bars at $z\sim0$ is
$\approx12$ per cent of the total stellar mass \citep[see also][]{WeiJogKho08},
which is a factor of three larger than what we find here. This discrepancy can
be explained, at least partially, by two factors. Firstly, the bar fraction assumed in the previous work
is 70 per cent, which is typically found in studies dedicated to estimate this value \citep[see e.g.][]{EskFroPog00},
whereas the fraction of disc galaxies with bars in this work is 42 per cent. Secondly, we have selected
our galaxy sample avoiding galaxies with low stellar mass. Thus, the stellar mass budget we find concerns
massive galaxies, and it is thus biased in favour of ellipticals, which can not contain the bars we are
addressing here. The fact that our bar fraction is low can also be explained by two factors. First,
as discussed in Sect. \ref{sec:method}, we miss most of the short bars, due to the limited spatial resolution
in SDSS images. Second, bar fraction seems to be larger in low-mass galaxies than in more massive galaxies
\citep{BarJogMar08}.

\begin{figure*}
   \centering
   \includegraphics[width=3cm,clip=true,keepaspectratio=true]{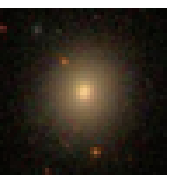}
   \includegraphics[width=3cm,clip=true,keepaspectratio=true]{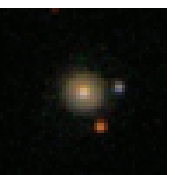}
   \includegraphics[width=3cm,clip=true,keepaspectratio=true]{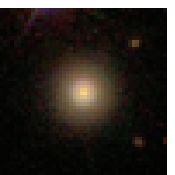}
   \includegraphics[width=3cm,clip=true,keepaspectratio=true]{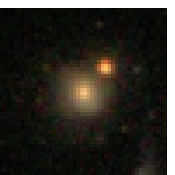}
   \includegraphics[width=3cm,clip=true,keepaspectratio=true]{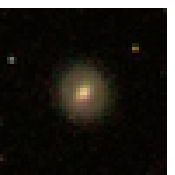}\\
   \includegraphics[width=3cm,clip=true,keepaspectratio=true]{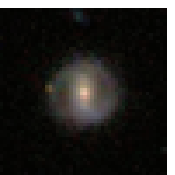}
   \includegraphics[width=3cm,clip=true,keepaspectratio=true]{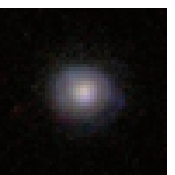}
   \includegraphics[width=3cm,clip=true,keepaspectratio=true]{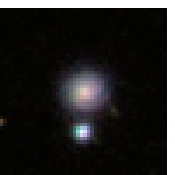}
   \includegraphics[width=3cm,clip=true,keepaspectratio=true]{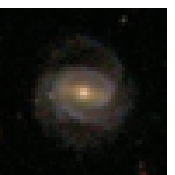}
   \includegraphics[width=3cm,clip=true,keepaspectratio=true]{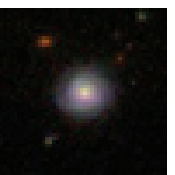}\\
   \caption{Examples of galaxies hosting classical bulges with D$_n{\rm(4000)}\geq1.7$ (top) and
D$_n{\rm(4000)}<1.7$ (bottom). It is evident that the latter show more often
spiral structure and blue discs.}
   \label{fig:ht}
\end{figure*}

\subsection{Classical bulges with star formation}

\begin{figure}
   \centering
   \includegraphics[keepaspectratio=true,width=8cm,clip=true]{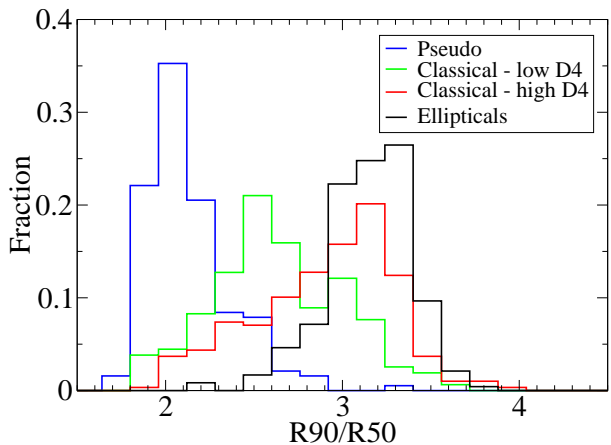}
   \caption{Distribution of concentration ($R90/R50$) for pseudo-bulges, classical bulges with
D$_n{\rm(4000)}<1.7$, classical bulges with D$_n{\rm(4000)}\geq1.7$, and elliptical galaxies.}
   \label{fig:conc_ht}
\end{figure}

One interesting finding from Figs. \ref{fig:d4col} and \ref{fig:d4mas} is that while
pseudo-bulges show almost always intense star formation, classical bulges can either be
quiescent or forming young stars. This indicates that galaxies with classical bulges can be further
divided according to their values of D$_n$(4000). We have thus separated classical bulges
in two different categories, namely those with D$_n{\rm(4000)}<1.7$, and those with
D$_n{\rm(4000)}\geq1.7$, and find that these categories have typically different
bulge-to-total ratios and stellar masses. Galaxies hosting classical bulges with D$_n{\rm(4000)}<1.7$
have median values of $B/T=0.26$ and mass equal to $0.8\times10^{10}~{\rm M}_\odot$,
whereas those with D$_n{\rm(4000)}\geq1.7$ have median values of $B/T=0.42$ and mass equal to
$1.8\times10^{10}~{\rm M}_\odot$.
This suggests that the former are dominated by early-type spirals whereas the latter
are mostly lenticular galaxies. To verify that, we have inspected randomly selected samples of these
two galaxy categories and concluded that, in fact, galaxies hosting classical bulges with
D$_n{\rm(4000)}<1.7$ show more often spiral structure and blue discs than galaxies hosting classical
bulges with D$_n{\rm(4000)}\geq1.7$ (see Fig. \ref{fig:ht}).
In addition, we show in Fig. \ref{fig:conc_ht} that these galaxies have markedly
different distributions of the concentration index $R90/R50$. Furthermore, galaxies with
pseudo-bulges also show a distribution of $R90/R50$ distinct from that of both
categories of galaxies with classical bulges. Elliptical galaxies, however, have a $R90/R50$
distribution more similar to that of galaxies with classical bulges and D$_n{\rm(4000)}\geq1.7$.
We thus see three peaks in the distribution of concentration. A related finding is reported in
\citet{BaiHar08}. As expected, the median S\'ersic index of star-forming classical bulges is 2.8,
thus lower than that of quiescent classical bulges, for which it is 3.6.

A plausible explanation for the existence of galaxies hosting classical bulges with conspicuous star
formation activity is the coexistence of a classical bulge and a pseudo-bulge in the same
galaxy. This would naturally occur if, after the formation of the classical bulge (e.g. via a
minor merger), the galaxy disc develops an instability (such as a bar) able to induce the
observed central star formation through processes of secular evolution.\footnote{Another
possibility is that star-forming classical bulges are those caught during a minor merger
that enhances central star formation and
will eventually help building the bulge. However, note that we have rejected
galaxies with disturbed morphologies (Sect. \ref{sec:sample}).}
Since pseudo-bulges are about an order of magnitude less massive than classical bulges
(see Table \ref{tab:masdis}), such {\em composite bulges} would have the structural signature
of a classical bulge, concurrent with the star formation activity of a pseudo-bulge, as observed.
The fact that the D$_n$(4000) values of star-forming classical bulges and pseudo-bulges are similar
(see Fig. \ref{fig:d4col}) is consistent with this picture.
Interestingly, we find that 34 per cent of galaxies with classical bulges have
D$_n{\rm(4000)}<1.7$ (50 per cent of which hosting bars), the remainder having D$_n{\rm(4000)}\geq1.7$
(with a fraction of barred galaxies of 40 per cent). Such excess of bars among
star-forming classical bulges is also generally consistent with the picture of a composite bulge.
In this context, it is worth noting that classical bulges with star formation lie between pseudo-bulges
and classical bulges without star formation in the density plot of Fig. \ref{fig:car99} (i.e. between
the lower dashed line and the solid line), and also have intermediate masses, as mentioned above (see also
Table \ref{tab:masdis}). They thus bridge pseudo-bulges and classical bulges without star formation
in the mass-size relation in Fig. \ref{fig:scale}. If such a picture is correct, then the Hubble
classification could be understood as mainly a sequence that
goes from elliptical galaxies successively to disc galaxies with classical bulges, composite bulges
and pseudo-bulges.

\subsection{Formation of bulges and ellipticals}

The results in Sect. \ref{sec:results} all corroborate previous ideas that bulges,
as identified through their
photometrical properties, are present in two main classes, with distinct structural properties,
indicating different formation processes. Those more similar to elliptical galaxies, namely classical
bulges, seem to be formed through violent processes, such as hierarchical clustering, whereas
those dissimilar bulges, namely pseudo-bulges, seem to be still under formation processes, more
likely as a result of their host galaxies' slow internal evolution, which involves disc instabilities
and angular momentum transfer.

These inferences are borne out by the fact that pseudo-bulges, when compared to their classical
counterparts, have currently intense star formation activity (Figs. \ref{fig:d4col} and
\ref{fig:d4mas}), are less concentrated at fixed bulge-to-total ratio
(Fig. \ref{fig:rehbt}), and occupy the same locus in the FP as discs (Fig. \ref{fig:kappa}).
Bulge S\'ersic index correlates with bulge-to-total ratio for classical bulges but not
for pseudo-bulges (Fig. \ref{fig:nbtmur}). In addition, pseudo-bulges follow a mass-size relation similar
to that followed by bars, and different from that followed by classical bulges
(Fig. \ref{fig:scale}). Furthermore, the discs (and bars) in galaxies
that host pseudo-bulges are not as different from those in galaxies that host classical
bulges as pseudo-bulges are different from classical bulges, in what concerns their
relative stellar masses (Sect. \ref{sec:discs}). This suggests
that disc and bar formation are similar in both galaxy classes, whereas bulge formation is distinct.

However, we have also found indications that the processes that lead to the formation of classical and
pseudo-bulges might happen concomitantly. In this context, the processes that lead to the formation of
a pseudo-bulge are dominant at one end, while those that lead to the formation of a classical bulge are
dominant at the other end, across the bulge population. This is suggested by the significant overlap
in the distributions of bulges with $n>2$ and $n<2$ in the Kormendy relation (Fig. \ref{fig:car99}), which
is in fact statistically more significant than the overlap in the distributions of ellipticals and
bulges altogether, as discussed in Sect. \ref{sec:pseudo}. We have also found classical bulges
with young stellar populations, typical of pseudo-bulges (Fig. \ref{fig:d4col}). Furthermore,
although classical and pseudo-bulges have significantly different slopes for their mass-size
relations, such relations join smoothly, with pseudo-bulges at the low-mass end, and classical
bulges at the high-mass end (Fig. \ref{fig:scale}).

Classical bulges seem to bear some structural similarity to ellipticals, since they follow the
same Kormendy relation (Fig. \ref{fig:car99}), and occupy a similar locus in the edge-on view of the FP
(Fig. \ref{fig:kappa}). However, they also follow offset mass-size relations (Fig. \ref{fig:scale}). This
could in principle challenge the idea that classical bulges are simply ellipticals that happen to be
surrounded by a disc. It should be noted, though, that this result concerns only massive ellipticals.
Low-luminosity ellipticals could, if surrounded by a disc, resemble classical bulges, but this can not be
evaluated with the present data. We have verified that
accounting for systematic effects could bring some of the
low-mass ellipticals closer to the relation followed by classical bulges. Our tests indicate that such
systematic effects should be restricted to about 10 per cent of the ellipticals. In the worst case
scenario that all ellipticals are affected by systematic effects, the offset between the mass-size
relations of classical bulges and ellipticals would still be present, but restricted to high-mass
systems. Thus, the result that high-mass bulges can not be considered as high-mass ellipticals surrounded
by discs seems to be robust. We note that in the face-on view of the FP (Fig. \ref{fig:kappa}),
pseudo-bulges, classical bulges and ellipticals occupy three significantly distinct loci, which
should provide clues to the different formation and evolution processes.

\section{Summary and conclusions}
\label{sec:sum}

We have determined several structural parameters for elliptical galaxies, bulges, discs and bars
through reliable multi-band image fitting of a representative sample of nearly 1000 local, massive
galaxies in the SDSS. We showed that
the Petrosian concentration index is a better proxy for bulge-to-total ratio
than the global S\'ersic index.

We showed that, while bulge S\'ersic index can be considered as a criterion to
distinguish pseudo-bulges from classical bulges, a more reliable, and physically motivated,
separation can be made using the Kormendy relation. While classical bulges follow the relation
set by ellipticals, pseudo-bulges do not, independently of their S\'ersic index. This shows
that classical bulges bear some structural similarity to elliptical galaxies, while pseudo-bulges
are structurally different.

Using D$_n$(4000), which is insensitive to dust attenuation, we demonstrated that pseudo-bulges
are currently undergoing intense star formation activity, and that virtually all pseudo-bulges
are located in the blue cloud of the colour-magnitude diagram. In contrast, most (but not
all) classical bulges are quiescent and populate the red sequence of the diagram. We presented
evidence of different formation mechanisms for classical and pseudo-bulges. Classical bulges
follow a correlation between S\'ersic index and bulge-to-total ratio, while pseudo-bulges do not.
The latter are less concentrated than the former at fixed bulge-to-total ratio.
The locus occupied by pseudo-bulges in the fundamental plane is different
from that of classical bulges, and it is the same as that of discs. Furthermore, pseudo-bulges
follow a mass-size relation different from that of classical bulges, and similar
to that of bars. Altogether, this
indicates that pseudo-bulges are formed through slow, non-violent processes, such as those expected
from disc instabilities, whereas classical bulges are formed through violent processes, such
as the merging of smaller units. We verified, however, a significant overlap in the properties
of classical and pseudo-bulges, which suggests that the different formation processes might happen
concomitantly, with different processes being dominant in different cases.

We found that classical bulges and ellipticals follow offset mass-size relations, suggesting
that high-mass bulges can not be considered as high-mass ellipticals that happen to be surrounded
by a disc.

We provided distributions and typical values (including uncertainties in individual measurements)
of all structural parameters obtained (with the exception of bar parameters, which will
be addressed in a separate paper).
We calculated the stellar mass content and distribution for each galaxy component,
and showed that, considering galaxies more massive than $10^{10}~{\rm M}_\odot$, in the local universe,
32 per cent of the total stellar mass is contained in ellipticals,
and the corresponding values for discs, bulges and bars are,
respectively, 36, 28 and 4 per cent. Classical bulges contain 25 per cent of the total stellar mass,
while pseudo-bulges contain 3 per cent. We also provided the corresponding luminosity and
number fractions. In particular, we find that approximately a third of disc galaxies hosts pseudo-bulges.
Such figures should be reproduced by successful models of galaxy formation and evolution.

\section*{Acknowledgments}
I am grateful to Guinevere Kauffmann for her support throughout this work and many useful
discussions. Her input is greatly appreciated.
It is a pleasure to thank Simone Weinmann for her help in producing Figure \ref{fig:conc_nyu}
and useful discussions, and
Jarle Brinchmann for producing the cutouts and detailed image headers of our parent sample.
I have benefitted from useful and stimulating discussions with Lia Athanassoula, Eugene Churazov, Peter Erwin,
David Fisher, Alister Graham, Tim Heckman, Jochen Liske, Daniele Pierini and Simon White throughout the
writing of this paper. I thank an anonymous referee for comments that helped to improve the paper.
DAG is supported by the Deutsche Forschungsgemeinschaft priority program 1177 (``Witnesses of Cosmic
History: Formation and evolution of galaxies, black holes and their environment''), and the Max Planck
Society.
The Sloan Digital Sky Survey (SDSS) is a joint project
of The University of Chicago, Fermilab, the Institute for Advanced
Study, the Japan Participation Group, The Johns Hopkins University,
Los Alamos National Laboratory, the Max-Planck Institute
for Astronomy (MPIA), the Max-Planck Institute for Astrophysics
(MPA), New Mexico State University, Princeton University, the
United States Naval Observatory, and the University of Washington.
Apache Point Observatory, site of the SDSS telescopes, is operated
by the Astrophysical Research Consortium (ARC). Funding
for the project has been provided by the Alfred P. Sloan Foundation,
the SDSS member institutions, the National Aeronautics and Space
Administration, the National Science Foundation, the US Department
of Energy, the Japanese Monbukagakusho, and the Max Planck
Society. The SDSS Web site is http://www.sdss.org/.

\bibliographystyle{mn2e}
\bibliography{../../gadotti_refs}

\appendix
\section{Fit sensitivity to input parameters}
\label{sec:app}

\begin{figure*}
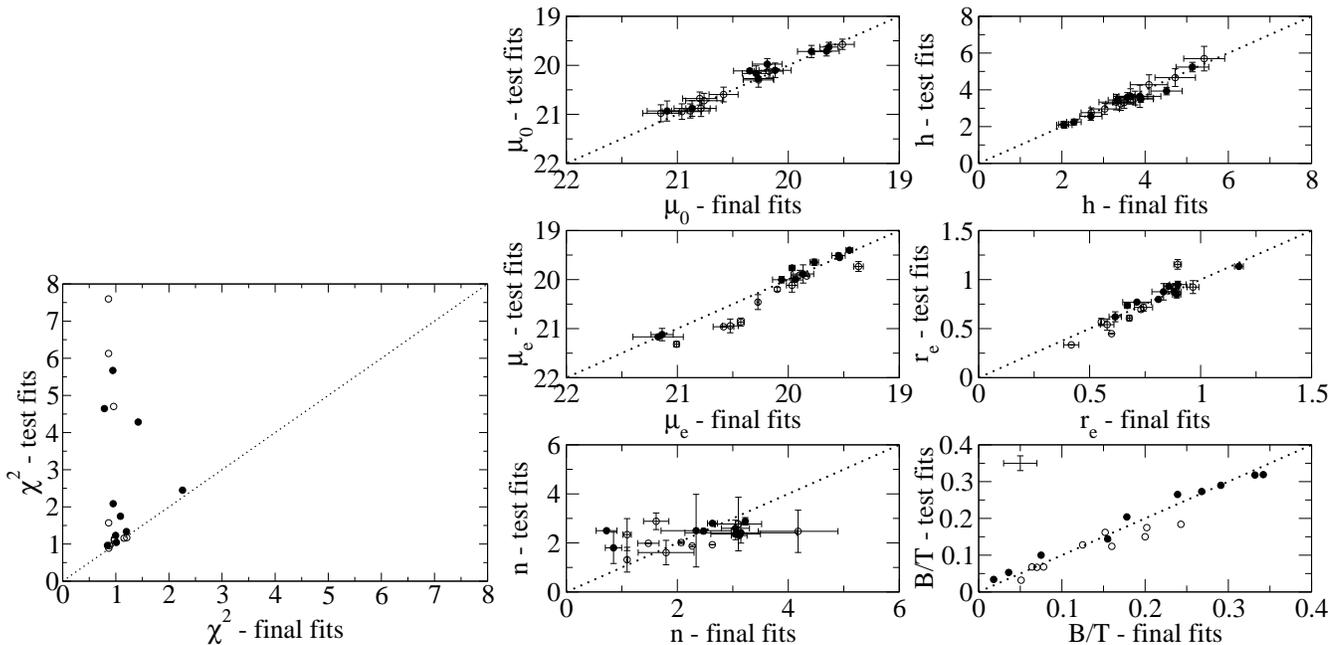

   \centering
   \includegraphics[keepaspectratio=true,width=6.5cm,clip=true]{newtest_chi.eps}
   \includegraphics[keepaspectratio=true,width=11cm,clip=true]{newtest.eps}
   \caption{Comparison between the results from test and final fits in twenty barred galaxies.
Filled circles refer to tests where the input parameters were multiplied by two, while empty
circles refer to tests where the input parameters were divided by two. Error bars are 1$\sigma$ uncertainties
estimated by the fitting code. For $B/T$, typical error bars are shown, calculated through error propagation
using the uncertainties in $\mu_e$, $r_e$ and $n$.}
   \label{fig:app}
\end{figure*}

As mentioned in Sect. \ref{sec:method}, image fitting often involves finding a $\chi^2$ global minimum in a very complex
multidimensional parameter space. This means that, in principle (and depending on the algorithm used),
fits can be very sensitive to the initial guesses
given to the fitting code to start the first iteration. To avoid erroneous results, we have opted to produce and check
each fit individually. Nevertheless, this alone does not mean that we find the most appropriate set of input parameters
in all cases.

In order to check how sensitive the results of our fits are to the set of initial guesses provided to {\sc budda},
we have performed the following tests. We have randomly selected twenty barred galaxies, and produced test fits, in
which the input values for $\mu_0$, $h$, $\mu_e$, $r_e$ and $n$ used in the corresponding final fits are
increased by a factor two (for ten of these galaxies) or decreased by a factor two (for the remaining ten galaxies).
The choice for barred galaxies assures that the results from these tests concern typically difficult fits.
Note also that a factor two is rather large, and that twenty galaxies correspond to 7 per cent of our barred galaxies.
Figure \ref{fig:app} shows the results from these test fits, as compared to those from the final fits.
Error bars are 1$\sigma$ uncertainties estimated by {\sc budda}. Its current version does not provide estimates
of the uncertainty in $B/T$. Thus, for $B/T$, typical error bars are shown,
calculated through error propagation using the uncertainties in $\mu_e$, $r_e$ and $n$, and assuming that the main
source of error in $B/T$ comes from determining the total bulge luminosity. Through similar tests with {\sc budda},
\citet{DurSulBut08} found comparable uncertainties in $B/T$ (albeit somewhat smaller).

One sees that, in most cases, the code converges to similar results, even for input parameters varying
by a factor four. In fact, most results are consistent within the estimated uncertainties.
All outliers come from the few fits that resulted in notably high $\chi^2$ values. Furthermore,
these results reinforce the findings in \citet{Gad08b}, by indicating that disc parameters are particularly robust,
and that $B/T$ is a more robust parameter than $n$, which can have large uncertainties compared to its dynamical
range. These results also argue in favour of our method to identify pseudo-bulges, at least for statistical studies
such as ours, since $\mu_e$ and $r_e$ are also more stable than $n$.

These results do not mean that any reasonable set of input parameters is appropriate. But they show that, by
creating input parameters through a careful inspection of each galaxy individually, we are able to
achieve robust results. In fact, as noted above, wrong fits often result in significantly higher $\chi^2$
values (see leftmost panel in Fig. \ref{fig:app}), turning them generally easy to recognise in our procedure.
Only 2.5 per cent of our final fits have $\chi^2>2$.

\label{lastpage}

\end{document}